# Should Benevolent Deception be Allowed in EHMI? A Mechanism Explanation Based on Game Theory

Linkun Liu, Jian Sun, and Ye Tian, *Senior Member, IEEE*

***Abstract*—The application of external human-machine interface (EHMI) on autonomous vehicles (AVs) facilitates information exchange. Existing research fails to consider the impact of the sequence of actions, as well as the effects of EHMI applications and deception, raising the question of whether benevolent, well-intentioned deception should be permitted (i.e., misleading statements that are intended to benefit both parties). We established a game theory based EHMI information disclosure framework for AVs in this study. In considering benevolent deception, this framework divided the decision-making process into three stages, respectively encompassing three key questions: whether to disclose, when to disclose, and what type of intention information to disclose. The results show that theoretical advantages of deception exist in certain cases when AV expects to maximize the safety of the interaction. In 40 out of 484 cases (8.3%), safety can be enhanced through successful deception. Those successful deceptions fall into two categories: 1) In 28 of these cases, the straight-going AV expected the left-turning human-driven vehicle (HV) to yield, while HV exhibited lower speed and higher acceleration; 2) In 12 of these cases, AV expected HV to proceed first, while HV exhibited higher speed and lower acceleration. We also conducted a VR-based driving simulation experiment, and the results confirmed our conclusion. Additionally, we found that when participants had low trust in the EHMI, its use negatively impacted interaction efficiency instead. This study serves as an exploratory behavioral mechanism study based on specific hypotheses for future EHMI design and ethical decision-making of autonomous driving system.**

## I. INTRODUCTION

THE great science fiction writer Isaac Asimov first introduced the famous Three Laws of Robotics in his short story "Runaround" in 1942, which established a set of moral and behavioral guidelines for robots [1]. These laws aim to direct robotic actions and enhance human welfare. However, behind the foolproof Three Laws of Robotics lie significant hidden risks.

Robots and humans may interpret the same information differently, which can lead to discrepancies in information transmission. Besides, humans are the weakest link in system security and have difficulty detecting deception [2]. Consequently, the issue of deception arises during the process of information disclosure, thereby making it challenging to ascertain the veracity of the information.

Autonomous Vehicles (AVs) emerge as a feasible solution to existing traffic challenges such as congestion, vehicle pollution, and traffic accidents [3]. Despite the rapid growth of autonomous driving system (ADS), conventional cars are still expected to account for roughly half by 2030 [4]. Prior to the large-scale deployment of AVs, increasing public trust remained a critical barrier to human-machine interaction [5].

Therefore, if we consider AVs as a type of robots, their rapid development inevitably presents issues related to information disclosure. The increasing frequency of interactions intensifies the need for feasible and reliable information disclosure.

Meanwhile, the External Human-Machine Interface (EHMI) is developing as a viable information disclosure channel. Driving information can be shown on the exterior in the form of a graph or text, allowing other road users to better understand the AV's status or purpose and promoting information exchange. Fig. 1 depicts the examples of EHMI. In low-speed urban environments, EHMI often relies on visual signals such as LED displays, light strips, or projections to convey vehicle intentions or warnings to pedestrians and cyclists [6], [7]. While on highways, EHMI may incorporate dynamic symbols or auditory signals to communicate lane changes, merging intentions, or emergency maneuvers to nearby drivers [8]. Moreover, the complexity of EHMI increases when considering mixed traffic environments. EHMI must be adaptive, context-aware, and capable of handling interaction targets with differing perceptual and cognitive capabilities [9], [10]. The communication effectiveness also depends on cultural factors, ambient lighting, weather conditions, and user trust levels [9]. We believe that EHMI for AV–HV interaction has already established a certain foundation and represents a promising direction for future development. It is thus worth conducting more forward-looking research in this area.

EHMI causes the direct disclosure of information, which may have a major impact on results. The establishment of trust is critical to the effectiveness of human-machine interactions [11]. Although deceitful behavior may provide people with short-term benefits, it eventually weakens the overall integrity

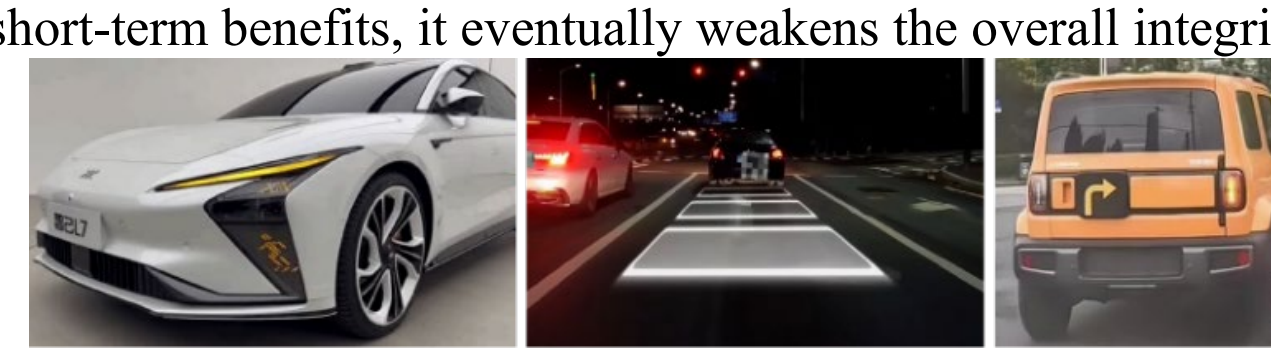

**Fig. 1.** Examples of EHMI.

The authors are all with the Department of Traffic Engineering and Key Laboratory of Road and Traffic Engineering, Ministry of Education, Tongji University. Shanghai, China. 201804. (E-mail: aliulinkun@tongji.edu.cn; sunjian@tongji.edu.cn; tianye@tongji.edu.cn)
Corresponding author: Ye Tian (tianye@tongji.edu.cn)

of the system. As a result, honesty is vital for achieving optimal social outcome [12].

As pointed out in [13], deception is not always inherently detrimental to the system in some cases. The focus is on finding optimal strategies in the presence of such deceptive vehicles [14]. **Currently, research involving deceptive vehicles typically assumes they are self-interested without considering the overall benefits.** While we unequivocally oppose selfish deception, the situation changes with the occurrence of white lies. A core requirement of deception is that a subject A must intend B to acquire a false belief [15]. And, if they are motivated by concern for others, they are “benevolent” [16]. “Benevolent deception” can thus be defined as the intentional act of causing another to hold a false belief, carried out with the primary aim of benefiting the deceived subject or both involved. This contrasts with general deception, which typically serves the deceiver's self-interest. For example, by employing benevolent deception, the recovery rate of patients can be increased. **Research has primarily focused on eliminating selfish deception while overlooking the potential benefits of using deception constructively** [17]**.** While in AV–HV interaction scenarios, a benevolent deception example from the HV's perspective can be defined as a case where the AV signals an intention to proceed first but deliberately yields instead, resulting in improved safety or efficiency for both parties. This raises the question of whether we should permit AV to disclose benevolent deception information?

In intersection scenario, human drivers typically attempt to "negotiate" through actions as signals such as accelerating to pass as quickly as possible. However, due to limitations in perception and reaction, they sometimes cannot avoid "vehicle face-off". Game theory can accurately describe how interactions between players reach equilibrium. The "vehicle face-off" represents an equilibrium where both players adopt the same strategy. **Existing research based on mechanism which does not consider EHMI information disclosure cannot avoid the impact of human intentions constantly changing during interactions.** Such as traditional game theory [18] or multi-agent reinforcement learning based on Markov chains [19]. In this case, while AVs can reach an agreement through continuous decision-making, it is unrealistic for humans to switch strategies multiple times in a short period based on actual conditions. In contrast, studies that do consider information disclosure assume that the disclosed information is truthful. Fully disclosing true information does not necessarily maximize the benefits for both. It is important to study whether AVs should disclose information, when to disclose information, and what intersection crossing strategies to disclose. **Existing methodology has difficulty to describe deceptional and truth-telling signals simultaneously.**

Deception is essentially a product of strategic information disclosure. Any act of deception can, in a broad sense, be described as a form of signaling game [20]. And the complex mechanisms of any deception can be abstracted and described through a signaling game model. If we treat the information disclosed by EHMI as signals, the signaling game can accurately describe EHMI's mechanisms and impacts.

Since the interaction object of autonomous vehicles is humans, who inevitably exhibit some special preferences in complex scenarios, this increases the decision-making difficulty for autonomous driving systems. Many studies have addressed human preferences in complex environments. For example, Crosato et al. [21] used Social Value Orientation (SVO) and Deep Reinforcement Learning (DRL) to generate decision-making strategies with different driving styles, addressing the issue of AVs coexisting with pedestrians in complex interaction environments. Wei et al. [22] emphasized the impact of human driver behavior and proposed a new concept of a risk-based corridor to limit the movement of AVs. However, these studies on human behavior mainly consider implicit information. And if there is information disclosure through EHMI, the possibility of deception inevitably arises. Therefore, in the presence of EHMI information disclosure, considering the complexity of the interaction scenarios and factors, it is necessary to establish a framework that describes the interaction process with humans.

Generally, most existing studies view deception as purely self-serving, overlooking its potential for mutual benefit. Moreover, current approaches often fail to incorporate EHMI-based signaling and struggle to effectively model both truthful and deceptive information. Therefore, there is a pressing need to develop a mechanistic explanation of both deceptive and truthful signals and to further investigate their potential impacts.

To overcome those challenges and explore the effectiveness of EHMI on conflict alleviation at intersections, this study is divided into three steps. Firstly, a signaling game model was utilized to analyze the interaction data of vehicles on real roads. Secondly, we calibrated the payoff parameters of the dynamic game model. Simulation analysis and empirical experiment were conducted based on the calibrated model to investigate the mechanism of EHMI. We established the optimal strategy for EHMI information disclosure. The goal is to maximize the payoff of the human-machine interaction at intersections.

The main contributions of this paper are as follows:

1) As far as the authors know, this is the first-ever work considering benevolent deception of EHMI.

2) This study is the first to employ the concept of signaling game to describe EHMI information disclosure. The process is deconstructed into three decision-making stages, where each subsequent stage must satisfy the conditions of the preceding one. It helps to clarify the path and logic of information disclosure.

3) This is an exploratory behavioral mechanism study based on specific hypotheses. The potential effects of long-term trust dynamics observed in the preliminary experiment offer a new perspective for future research and design of signaling mechanisms.

The remaining sections of this paper are organized as follows: Section 2 reviews the decision-making modeling method. Section 3 describes the methodology, including model formulation and calibration. Section 4 introduces the numerical experiment and the analysis of the data. Section 5 analyzes the impact of EHMI strategies and demonstrates the superiority of successful benevolent deception through experiments. Section 6 concludes the paper and provides future perspectives.

## II. Literature Review

In Europe, 40% of traffic accidents were intersection-related crashes [23]. Additionally, both Waymo and human drivers are reported to have difficulties to master unprotected left turns [24]. Vehicle face-off is the primary factor contributing to the danger in unprotected left-turn scenarios. As they approach the conflict zone, drivers often repeatedly make the same synchronous decisions: to accelerate or decelerate [24]. Because they sway between yielding and not yielding, both vehicles face multiple reasonable choices, leading to decreased efficiency and even collisions [25]. Therefore, effective interaction strategies in unprotected left-turn scenarios are crucial for ensuring safety.

In recent years, various decision-making algorithms in the intelligent system have been proposed by researchers [26]. In the context of vehicle decision-making modeling, there exist two primary categories of approaches: learning-based methods and mechanism-based methods.

Learning-based methods apply machine learning and deep learning techniques to extract optimal decision strategies by assimilating and analyzing substantial amounts of data. Zhang et al. [27] developed a reinforcement learning-based prediction framework for vehicle trajectory planning. Rolando et al. [28] developed an integrated safety-enhanced framework for AV that combines reinforcement learning and model predictive control. Qi et al. [29] modeled vehicle turning behavior at conflicting areas of intersections based on deep learning. Ma et al. [30] used graph to describe complex interactions.

Mechanism-based methods focus on individual decision-making for multiple participants, enabling a comprehensive and customized evaluation of safety, efficiency, and comfort from a mechanistic perspective. And game theory is a widely adopted approach in this regard.

AVs can disclose information explicitly through EHMI, implicitly through dynamic movements, or through a combination of both methods [31]. When EHMI is not considered, learning-based methods can effectively capture implicit information. However, when studying the mechanisms of EHMI, decision-making models that rely exclusively on learning methods are not suitable, which necessitate a substantial quantity of training data and exhibit limited interpretability. Mechanism-based methods abstract complex factors into easily interpretable models and do not require extensive real-world data, making them ideal for focusing on the mechanisms of EHMI. As discussed previously, game theory is well-suited for investigating vehicle interaction behaviors, and signaling games are particularly suitable for examining the impact of EHMI in these interactions. Signaling games also need to analyze the impact of the authenticity of disclosed information, making them particularly suitable for studying the benevolent deception of EHMI. This is a capability that other methods find difficult to match.

Many studies have approached the control of vehicle motion from the perspective of games. These models aim to govern the movement of vehicles at each step. Shao et al. [32] demonstrated that models based on signaling games outperform traditional lane-changing models using simulation. Li et al. [33] introduced a multi-stage repeated game and a probability model to describe the driver's tendency to accelerate in the interaction at a signal-free intersection. Hang et al. [34] established a new decision-making framework using differential game theory, constructing a collision risk assessment model to reduce computational complexity.

Many studies have also validated the proposed game algorithms by calibrating and verifying them using real-world datasets. Shao et al. [32] performed simulation validation after calibration using the NGSIM dataset. Hang et al. [18] identified and extracted feature patterns and characteristics of three driving styles from real-driving data. Talebpour et al. [35] proposed a sequential move structure based on the data from the perception system of autonomous vehicles.

When considering deception, as the study mentioned previously, a simple autonomous intersection management system, controlled by a hypothetical central roadside unit (RSU), was introduced [11]. The RSU receives information from vehicles. Selfish and deceptive vehicles can broadcast false information to the RSU to pass through first. The introduction of deceptive vehicles has been demonstrated to have a detrimental impact on the intersection's capacity, according to simulation. However, in certain specific situations, such as when the flow rates of each lane differ, the addition of deceptive vehicles occasionally results in an improvement in traffic efficiency. The positive impact of selfish, deceptive vehicles in reducing the cumulative delay of subsequent vehicles may outweigh their negative impact on the overall system, leading to an overall improvement. While this study offers a novel perspective on understanding deceptive behavior in traffic systems, it does not provide a precise explanation of the underlying mechanisms, as any unexpected increase in overall benefit is simply a byproduct of specific scenarios.

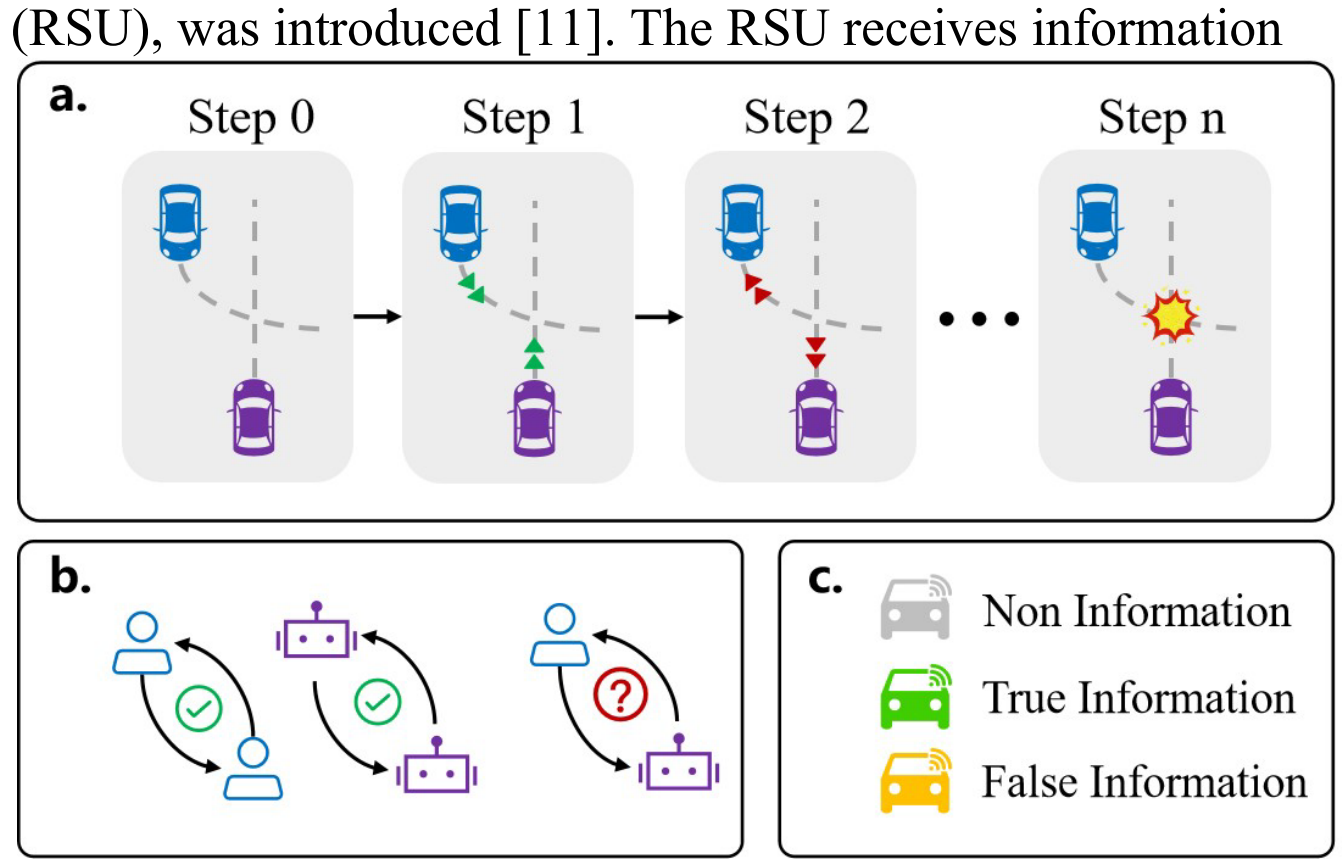

**Fig. 2.** Research gaps. a), Repeated games lead to strategy shifts. b), Limited research on diverse types of information disclosure. c), The authenticity of the disclosed information.

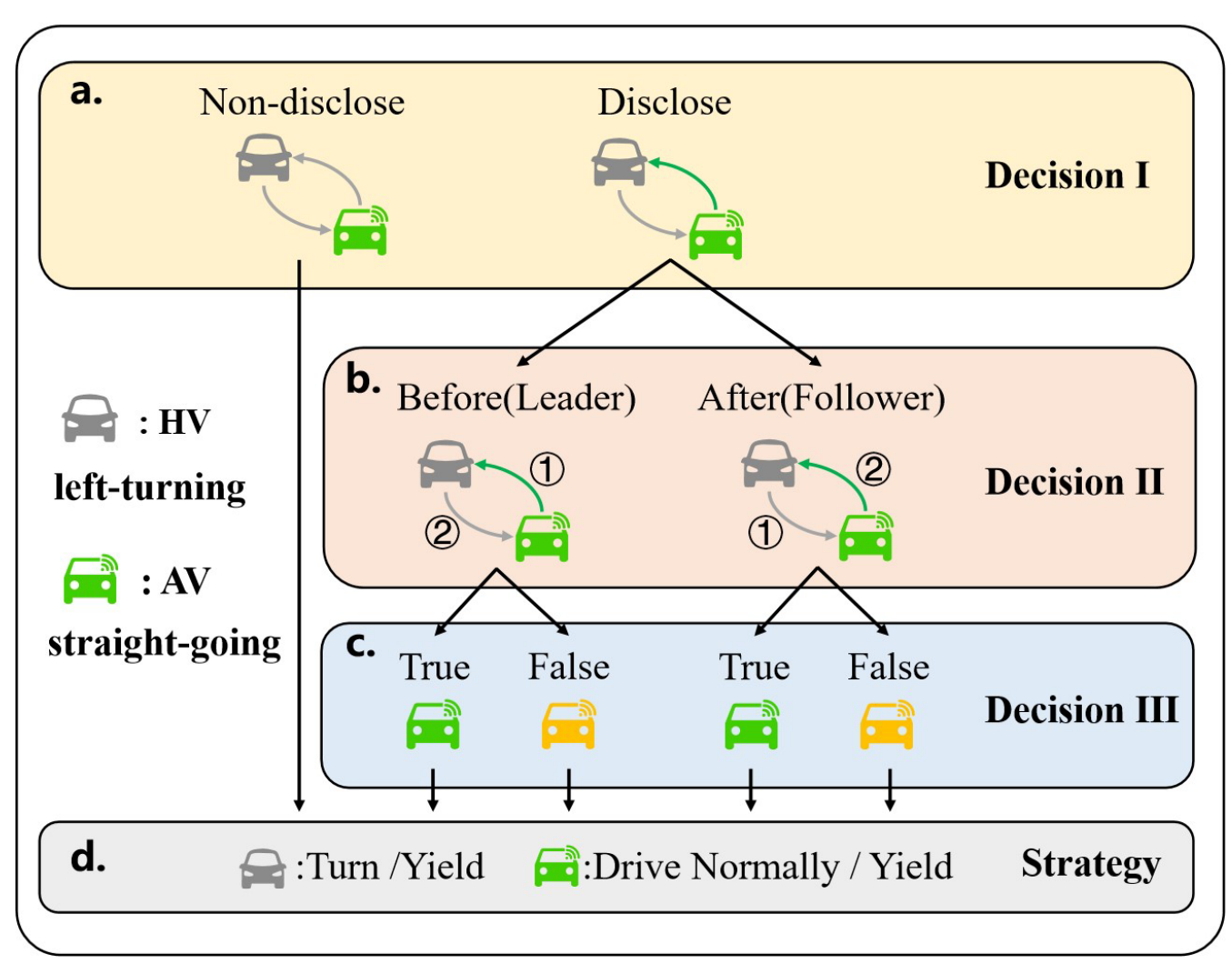

**Fig. 3.** The decision-making stages for interactive behavior.

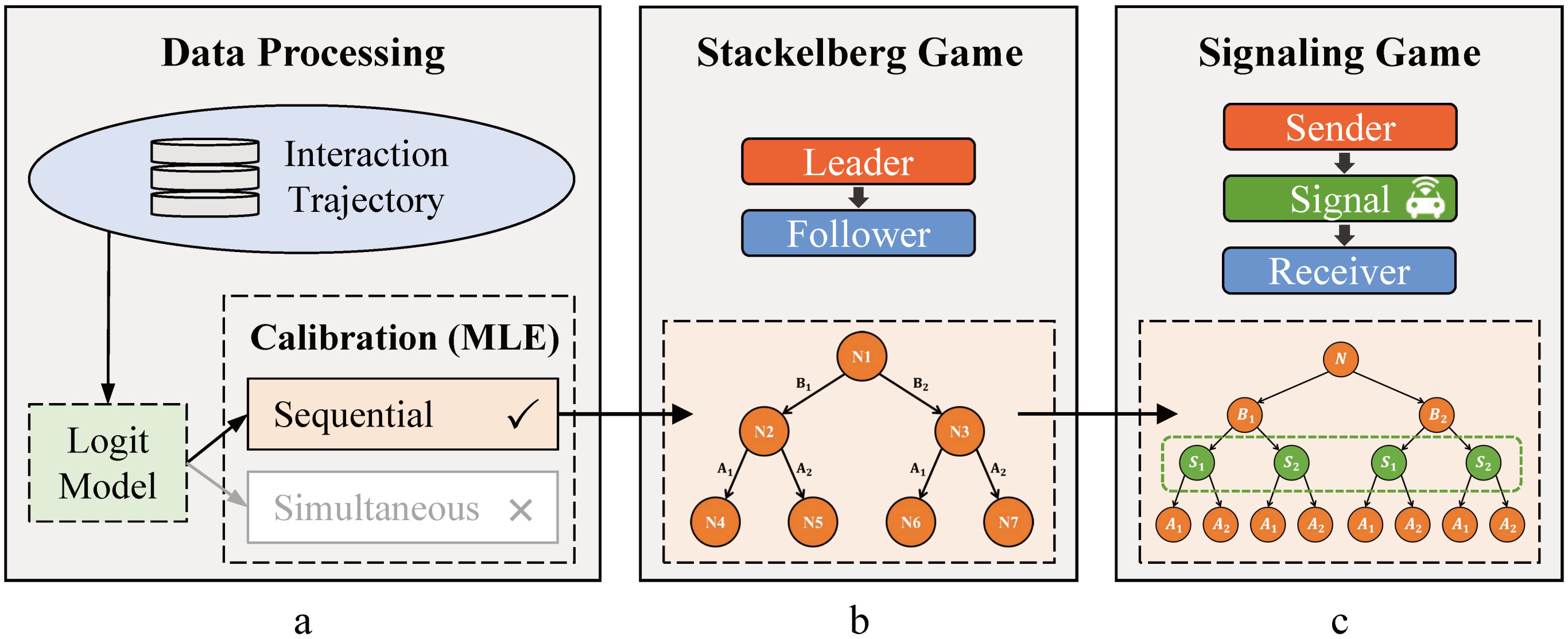

**Fig. 4.** The research flowchart. (a): Model calibration based on real-world trajectories. (b): Selection of the sequential-move game as the baseline. (c): Extension to a signaling game incorporating EHMI strategies.

**Current research has three main limitations**: First, existing interaction algorithms do not account for the sequential impact of information exchange and actions by both players. This can lead to strategy switching due to information asymmetry, creating safety risks (Fig. 2(a)). Second, there is less research on diverse types of information disclosure than on interactions in typical "human-human" and "machine-machine" situations (Fig. 2(b)). Third, as shown in Fig. 2(c), existing research does not consider the impact of EHMI applications, let alone benevolent deception.

Moreover, most research on EHMI is experimental and focuses primarily on vehicle-pedestrian interactions. There is a lack of theoretical modeling that considers the impact of EHMI applications on AV-HV interactions.

## III. Methodology

To address the limitations of the research, we developed a decision-making framework for unprotected left turns in human-machine interactions under EHMI applications based on signaling game. This framework incorporates elements of benevolent deception for further analysis.

In an unprotected left-turn scenario, the human-machine interaction involves a human driver making a left turn and an AV equipped with EHMI going straight. This specific case was intentionally selected to encourage interactional behavior that aligns with the intentions of AVs. This setup takes a real-world scenario into consideration, where vehicles proceeding straight typically have higher priority during the interaction.

The game strategies include two parts: 1) the AV drives normally or yields, and 2) the HV turns or yields. When considering the EHMI in AV, the decision-making process can be divided to three stages, as illustrated in Fig. 3.

The first decision (Fig. 3(a)) is whether the AV should disclose its intention. If it chooses not to disclose its intention, information disclosure won't be considered in the subsequent game, without the need for making the 2nd and 3rd decisions. If the AV chooses to disclose its intention, the following decision-making stages occur.

The second decision (Fig. 3(b)) is when the AV should disclose its intention. Here, "when" does not refer to a specific time but rather to the sequence of actions. In the dynamic game, AV is regarded as the leader if it discloses its intention before the opponent acts; if not, it is regarded as the follower.

The third decision (Fig. 3(c)) is what intention the AV should disclose, i.e., whether it should disclose the true intention or provide a false intention.

In these stages, different decision processes lead to different results (Fig. 3(d)). It is crucial to investigate whether AVs should disclose information, when to disclose it, and what strategies to disclose.

Notably, the EHMI disclosure strategy here is based on the expected strategy derived by AVs for the current interaction scenario. In our study, it is set to maximize the total payoff considering safety and comfort for both players. In the third decision stage, a benevolent deception strategy is considered successful if the final actual interaction strategy aligns with the AV's expected strategy.

As illustrated in Fig. 4, the research consists of three main steps. First, we calibrated the payoff functions of the game models under different action sequences (sequential-move and simultaneous-move) based on interaction trajectories in the real world. The logit model and maximum likelihood estimation were employed for model calibration. Next, the sequential-move game (i.e., the Stackelberg Game) was selected to represent the baseline scenario (i.e., the scenario without EHMI information disclosure) since such a game format better aligns with the actual action sequence as compared to simultaneous-move. Then, we extend the baseline model into a signaling game by incorporating the impact of two different EHMI strategies (rush or yield) as signals, where leader and follower correspond to sender and receiver, respectively. This inevitably introduces the possibility of deception, where disclosed information may not always align with actual behavior. The focus will be on benevolent deception, exploring its implications and discussing its role in interactions.

*A. Model Formulation*

1) **Player and Strategy Sets**

Fig. 5 provides an illustrative diagram of the unprotected left-turn conflict scenario. The present study examines the dynamic interaction between two distinct drivers, designated as A and B. The interaction is initiated upon mutual recognition of each other's actions when they are both simultaneously at the intersection. According to the expected sequence for passing through the conflict zone, each vehicle has two strategies. For the left-turning vehicle A, the strategies are "1) turn and 2) yield", and for the straight-going vehicle B, the strategies are "1) drive normally and 2) yield". TABLE I presents the payoffs ($U_{ij}$) for each vehicle under the four strategy sets ($o_{ij}$). $i$ and $j$ represent the strategies chosen by vehicle $A$ and $B$, respectively.

The strategy sets simply represent the parties' prospective intents based on their initial interaction state. However, this may not always correspond with the actual action methods used throughout the ensuing engagement.

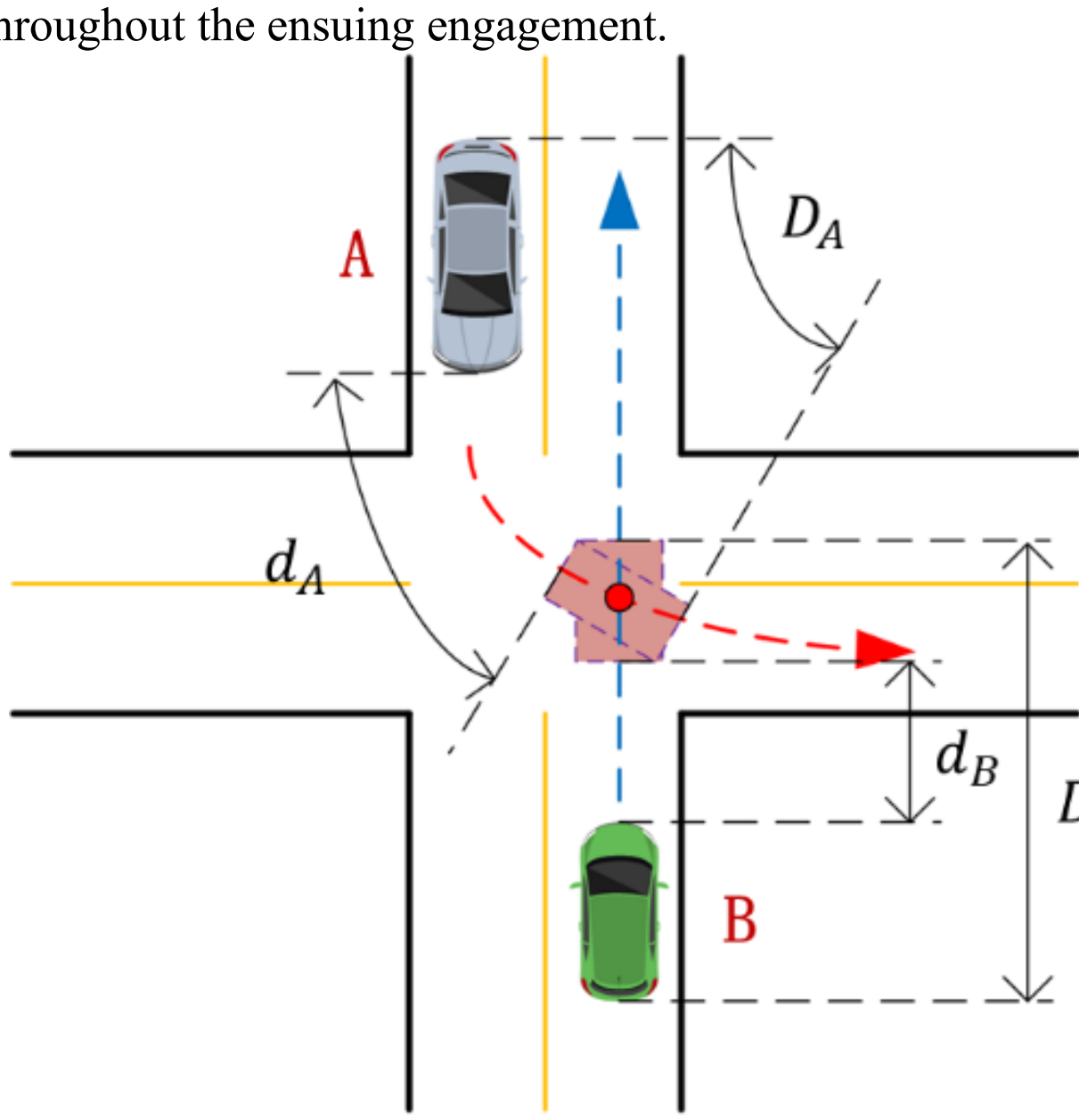

**Fig. 5.** The layout of the unprotected left turn scenario.

TABLE I
PAYOFF MATRIX OF THE GAME

| Left-turning Vehicle A | Straight-going Vehicle B | |
|---|---|---|
| | $B_1$ | $B_2$ |
| $A_1$ | $U_{11}^A, U_{11}^B$ | $U_{12}^A, U_{12}^B$ |
| $A_2$ | $U_{21}^A, U_{21}^B$ | $U_{22}^A, U_{22}^B$ |

2) **Payoff Functions**

According to experimental studies, accident risk and comfort are the most important elements affecting drivers' decisions in unprotected left turns [36]. Our reward function incorporates both risk and comfort through acceleration. We consider using indirect indicators, rather than directly incorporating direct indicators (i.e., TTC, PET). This is because the ultimate impact of these indicators is reflected in acceleration.

Due to the inability of external observers to capture all the factors that may affect players' decisions, it is necessary to consider error terms when formulating the game payoff functions to obtain more realistic results. The payoff is defined as a function of the deterministic component and the error term $\varepsilon$ in (1). The deterministic component is a function of the parameter vector $\theta$ and the observable factor vector $X$, which influences players' decisions. (Only the general form is presented here, while specific parameters and factors are introduced later.)

$$U = v(\theta, X) + \varepsilon \tag{1}$$

The decision to employ acceleration to measure payoffs is intended to account for implicit intent information while minimizing the impact of various dimensions. The model adopts one vehicle's perspective, relying only on vehicles' observable driving actions, as the interacting driver's' internal physiological or psychological indicators are unobservable. Additionally, to consider the impact of interactions, we introduce a parameter $a_c^i$, representing the maximum acceleration for collision avoidance. $i$ represents the player, referring to either vehicle A or B.

First, we use the following kinematic formulas to calculate the time for both to reach the conflict zone and completely traverse it. For the initial moment of game interaction, the vehicle $i$ has a velocity of $v_i$ and an acceleration of $a_i$. The distance traveled to reach the conflict zone is denoted as $d_i$, while the distance traveled to completely pass through the conflict zone is represented by $D_i$, as illustrated in Fig. 5. "Reaching the conflict zone" refers to the point at which the front of the vehicle contacts the conflict zone, while "completely passing through the conflict zone" indicates when the rear of the vehicle has departed from the conflict zone. Shown as (2) and (3).

$$t_0^i = sign(a_i) \times \sqrt{\left(\left(\frac{v_i}{a_i}\right)^2 + \frac{2d_i}{a_i}\right)} - \frac{v_i}{a_i} \tag{2}$$

$$t_1^i = sign(a_i) \times \sqrt{\left(\left(\frac{v_i}{a_i}\right)^2 + \frac{2D_i}{a_i}\right)} - \frac{v_i}{a_i} \tag{3}$$

The time taken to reach the conflict zone is denoted as $t_0^i$, while the time taken to completely pass through the conflict zone is denoted as $t_1^i$.

$$a_c^A = \frac{2 \times (d_B - v_B t_1^B)}{(t_1^B)^2} \tag{4}$$

$$a_c^B = \frac{2 \times (d_A - v_A t_1^A)}{(t_1^A)^2} \tag{5}$$

Thus, in (4) and (5), the values of $a_c^A$ and $a_c^B$ represent the maximum acceleration that vehicle A and vehicle B can achieve, respectively, to avoid collisions.

Regarding the composition of the payoff functions, both players first need to consider the impact of acceleration. If the opponent's strategy involves preemptively passing through the conflict area, its own $a_c^i$ must be considered in the payoff. Conversely, $a_c^i$ is not considered when the opponent chooses to yield. The payoff functions for player A's four strategies are denoted as Eq. (6, 7, 8, 9), while those for B's strategies are denoted as Eq. (10, 11, 12, 13). $\alpha, \beta$ are the parameters to be calibrated.

$$U_{11}^A = \alpha_{11}^0 + \alpha_{11}^1 a_A + \alpha_{11}^2 a_c^A + \varepsilon_{11}^A \tag{6}$$
$$U_{12}^A = \alpha_{12}^0 + \alpha_{12}^1 a_A + \varepsilon_{12}^A \tag{7}$$
$$U_{21}^A = \alpha_{21}^0 + \alpha_{21}^1 a_A + \alpha_{21}^2 a_c^A + \varepsilon_{21}^A \tag{8}$$
$$U_{22}^A = \alpha_{22}^0 + \alpha_{22}^1 a_A + \varepsilon_{22}^A \tag{9}$$

$$U_{11}^B = \beta_{11}^0 + \beta_{11}^1 a_B + \beta_{11}^2 a_c^B + \varepsilon_{11}^B \tag{10}$$
$$U_{12}^B = \beta_{12}^0 + \beta_{12}^1 a_B + \beta_{12}^2 a_c^B + \varepsilon_{11}^B \tag{11}$$
$$U_{21}^B = \beta_{21}^0 + \beta_{21}^1 a_B + \varepsilon_{21}^B \tag{12}$$
$$U_{22}^B = \beta_{22}^0 + \beta_{22}^1 a_B + \varepsilon_{22}^B \tag{13}$$

*B. Method of Calibration*

Based on game theory, calibration is performed by estimating the values that are most consistent with the observed data. The calibration method is based on the inverse analysis of the interactive behavior between the two players.

1) **Probability of Nash Equilibrium Point**

There are two sequences of actions for the game players: simultaneous moves and sequential moves. To better analyze the process, this study examines the Nash equilibrium in both forms of games.

a) Simultaneous-Move Games:

Let $o_{11}$ represent the selection of strategy $A_1B_1$ by the players. The necessary condition for $o_{11}$ being a Nash equilibrium is denoted as (14). However, when (15) is satisfied, both $o_{11}$ and $o_{22}$ will be the Nash equilibrium state of the game.

$$U_{11}^A \geq U_{21}^A, U_{11}^B \geq U_{12}^B \tag{14}$$

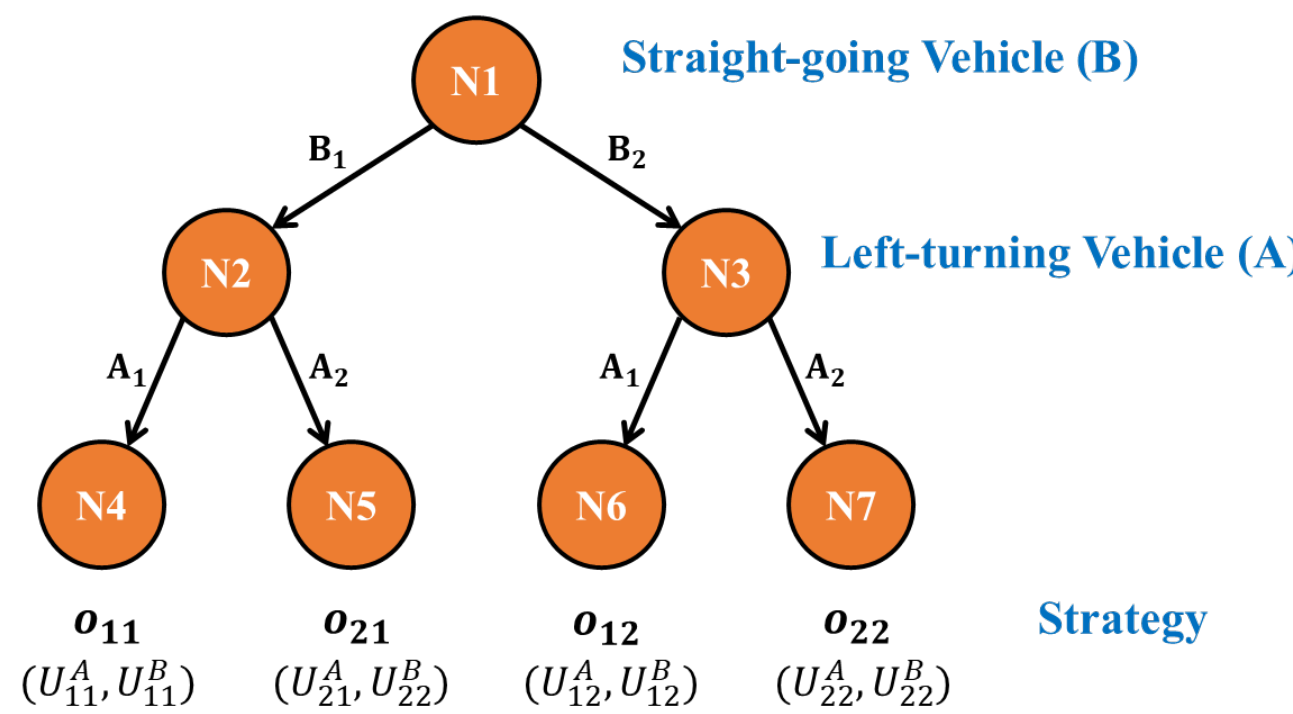

**Fig. 6.** Tree Diagram of the Stackelberg Game.

$$U_{11}^A \geq U_{21}^A, U_{11}^B \geq U_{12}^B, U_{22}^A \geq U_{12}^A, U_{22}^B \geq U_{21}^B \tag{15}$$

In this study, it is assumed that when there are two Nash equilibriums, they are chosen with probabilities $\delta$ and (1-$\delta$), respectively. The value of $\delta$ is obtained from the proportion in the data. Therefore, in this game, the probability of strategy $o_{11}$ being designated as an equilibrium point is shown in (16). The probabilities of the remaining strategies can be calculated using the same method.

$$\begin{aligned} P_{o11} = & P[U_{11}^A \geq U_{21}^A, U_{11}^B \geq U_{12}^B, U_{12}^A \geq U_{22}^A, U_{21}^B \geq U_{22}^B] + \\ & P[U_{11}^A \geq U_{21}^A, U_{11}^B \geq U_{12}^B, U_{12}^A \geq U_{22}^A, U_{22}^B \geq U_{21}^B] + \\ & P[U_{11}^A \geq U_{21}^A, U_{11}^B \geq U_{12}^B, U_{22}^A \geq U_{12}^A, U_{21}^B \geq U_{22}^B] + \\ & \delta P[U_{11}^A \geq U_{21}^A, U_{11}^B \geq U_{12}^B, U_{22}^A \geq U_{12}^A, U_{22}^B \geq U_{21}^B] \end{aligned} \tag{16}$$

b) Sequential-Move (Stackelberg) Games:

The determination of Nash equilibrium states in this context requires the application of backward induction. The tree diagram of the Stackelberg game is presented in Fig. 6, if the straight-going vehicle B acts first.

Assuming $o_{11}$ is a Nash equilibrium state, from N4 to N2 to N1, (17) and (18) should be satisfied, respectively:

$$U_{11}^A \geq U_{21}^A \tag{17}$$

$$U_{12}^A > U_{22}^A, U_{11}^B > U_{12}^B \text{ or } U_{22}^A > U_{12}^A, U_{11}^B > U_{22}^B \tag{18}$$

Therefore, in this game, the probability is given by (19).

$$\begin{aligned} P_{o_{11}} = & P[U_{11}^A > U_{21}^A, U_{12}^A > U_{22}^A, U_{11}^B > U_{12}^B] \\ & + P[U_{11}^A > U_{21}^A, U_{22}^A > U_{12}^A, U_{11}^B > U_{22}^B] \end{aligned} \tag{19}$$

2) **Logit Model**

In accordance with the logit model, the probability of an option being selected is contingent upon its utility. Consequently, the probabilities of the four strategy sets occurring in game $n$ can be calculated as indicated in (20). $k$ represents player A or B. $i$ and $j$ represent different strategies. $v$ represents the deterministic part of the payoff function, excluding the error term.

$$P\left(U_{i|n}^{k} \geq U_{j|n}^{k}\right) = \frac{e^{v_{i|n}^{k}}}{e^{v_{i|n}^{k}} + e^{v_{j|n}^{k}}} \tag{20}$$

3) **Maximum Likelihood Estimate (MLE)**

Based on the Logit model, parameter estimation is performed using MLE by maximizing the logarithm of the likelihood function. We chose MLE due to its interpretability, computational efficiency, and its suitability for the logit-based decision framework used in our game model. The likelihood function is shown in (21):

$$L(\theta) = \prod_{n}^{N} \prod_{s} P_{o_s|n}^{y(o_s|n)} \tag{21}$$

where $\theta$ represents the parameter vector of the payoff function, while $L(\theta)$ denotes the likelihood function. We define a binary indicator, $y(o_s \mid n)$, to indicate whether strategy $s$ is chosen ($y(o_s \mid n) = 1$) or not chosen ($y(o_s \mid n) = 0$) in game $n$.

Besides, while MLE has been widely applied in behavioral modeling, its limitations—such as the inability to capture unobserved heterogeneity or dynamic decision-making patterns of complex traffic data—can't be ignored [37].

## IV. Numerical Experiment

### A. Data Processing

For a case study, we selected HD aerial video from 4:00 PM to 5:30 PM on a weekday at the Jianhe-Xianxia Road intersection in Shanghai. From this footage, we extracted trajectory data for 484 unprotected left-turn interactions between two vehicles (The extracted data, the calibration code, and results of the cross-validation can be found at the following Git link: https://github.com/liulinkun6677/MLE-Calibration-for-Game-Theory.git). Fig. 7 shows the extracted trajectories.

### B. Distinction of Intentional Strategy

The extracted trajectories only reflect the actual interaction between the two vehicles intuitively and do not fully capture the initial intent when they started interacting. Taking the strategy $A_1B_1$ as an example, this implies that the straight-going vehicle intends to proceed normally, while the left-turning vehicle aims to aggressively turn left. However, during the subsequent interaction process, both vehicles continuously update their strategies. As a result, the trajectory data reflects either the left-turning vehicle turning first while the straight-going vehicle yields, or the left-turning vehicle yielding while the straight-going vehicle proceeds normally.

Therefore, we chose to differentiate intention strategies based on changes in speed during interaction. The velocity variation over each frame is depicted in Fig. 8. The velocity of the left-turning vehicle is shown in red, while that of the straight-going vehicle is shown in blue. The letters a, b, c, d respectively correspond to the intention strategies: $A_1B_1, A_1B_2, A_2B_1, A_2B_2$.

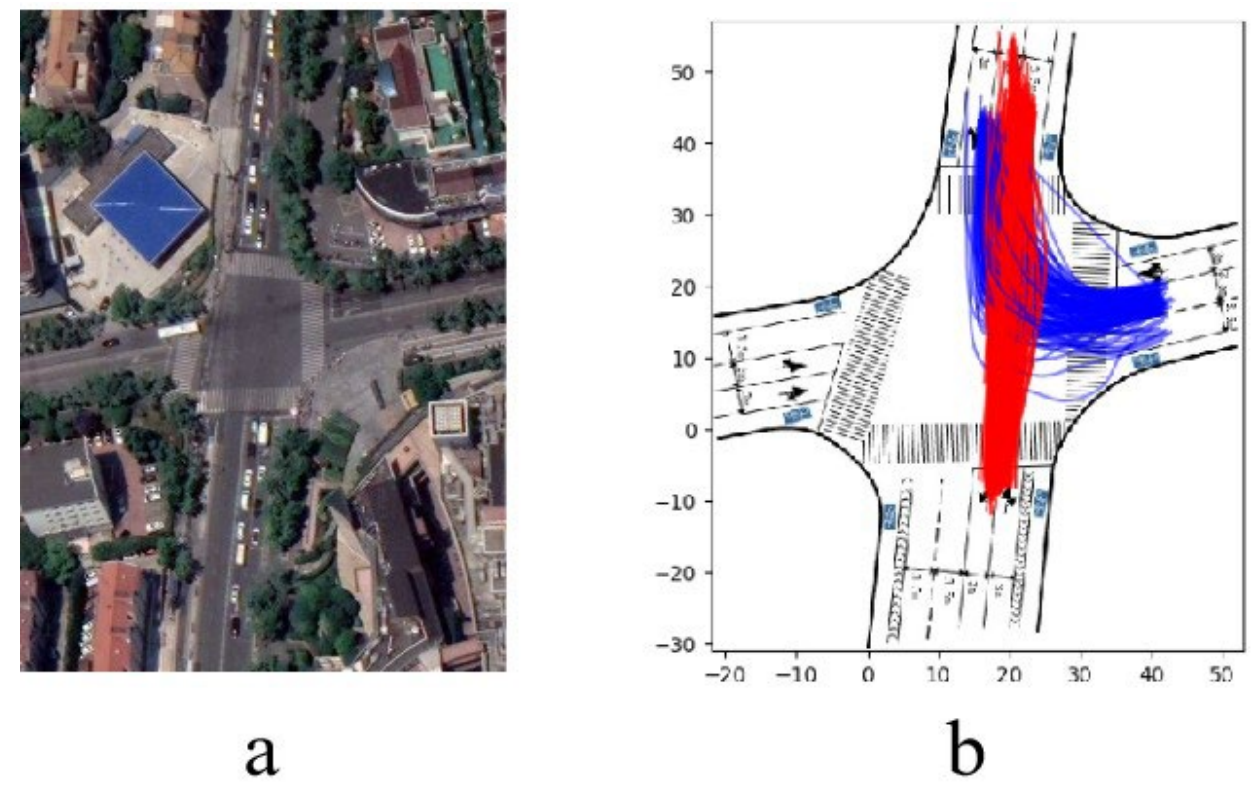

**Fig. 7.** The observed trajectories of the intersection.

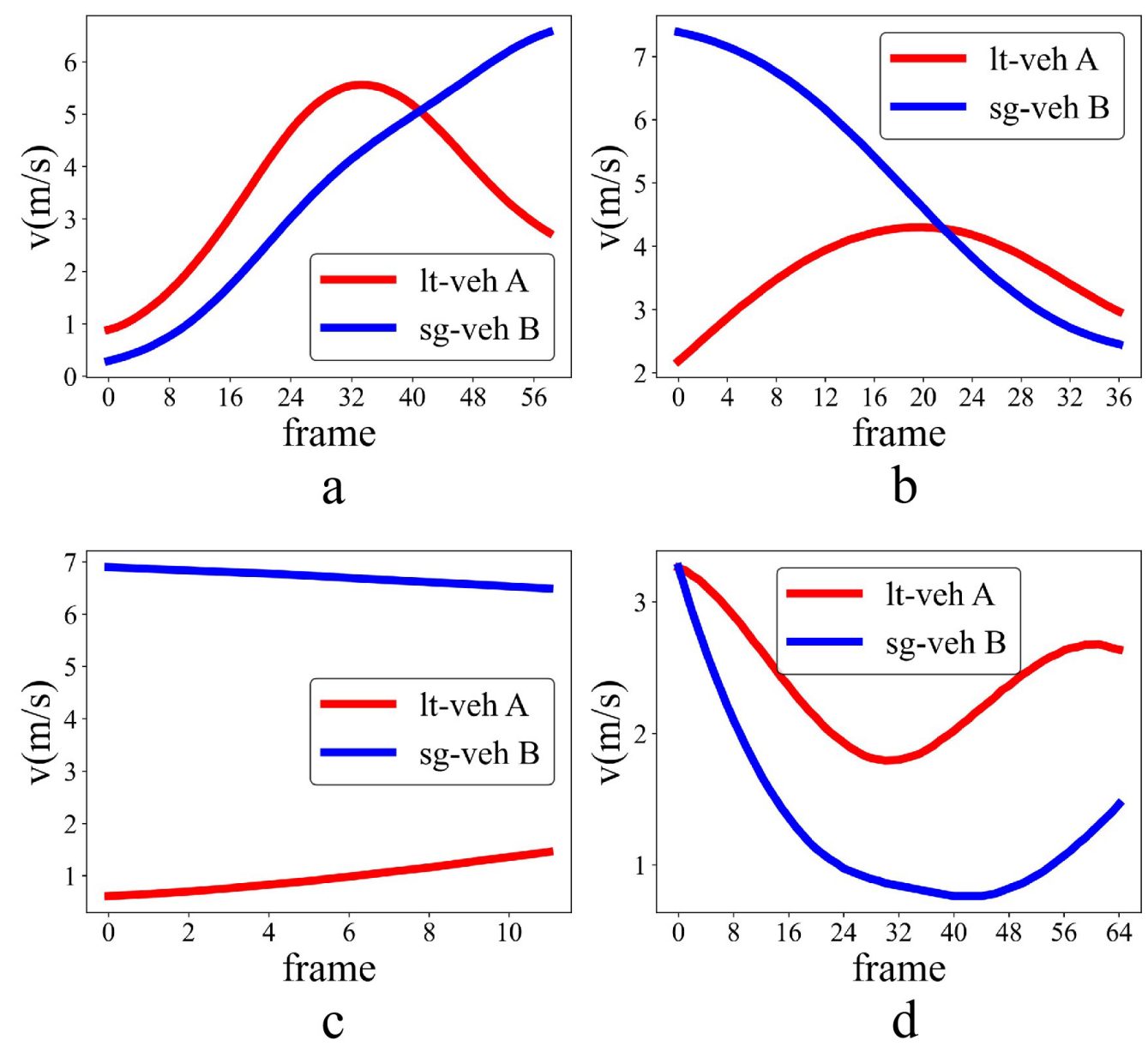

**Fig. 8.** Velocity variation of left-turning vehicle (lt-veh) and straight-going vehicle (sg-veh) over each frame for different intention strategies.

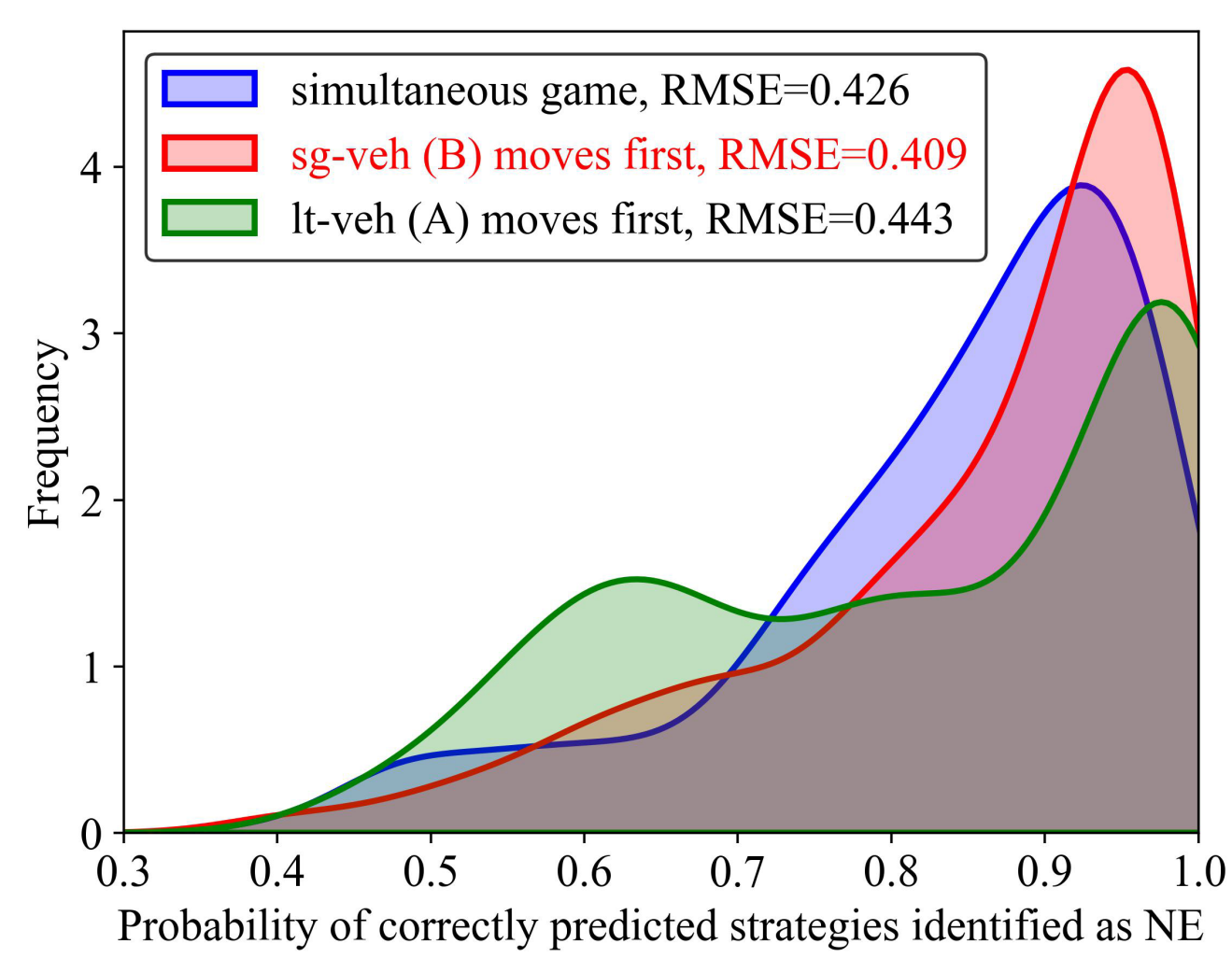

**Fig. 9.** The probability distribution of correctly predicted strategies that were identified as Nash equilibrium states.

TABLE II
CALIBRATION RESULTS

| Parameter | Results | Parameter | Results |
|---|---|---|---|
| $\alpha_{11}^0$ | 0.954 | $\beta_{11}^0$ | 1.277 |
| $\alpha_{11}^1$ | 5.107 | $\beta_{11}^1$ | 3.276 |
| $\alpha_{11}^2$ | 1.273 | $\beta_{11}^2$ | 1.505 |
| $\alpha_{12}^0$ | 2.440 | $\beta_{12}^0$ | 2.723 |
| $\alpha_{12}^1$ | 2.950 | $\beta_{12}^1$ | 0.724 |
| $\alpha_{21}^0$ | 3.359 | $\beta_{12}^2$ | 2.495 |
| $\alpha_{21}^1$ | 1.174 | $\beta_{21}^0$ | 3.435 |
| $\alpha_{21}^2$ | 4.748 | $\beta_{21}^1$ | 2.969 |
| $\alpha_{22}^0$ | 1.245 | $\beta_{22}^0$ | 0.565 |
| $\alpha_{22}^1$ | -1.232 | $\beta_{22}^1$ | 1.030 |

### C. Model Calibration and Validation

Calibration of the model parameters was conducted for both sequential and simultaneous-move games. To validate the model, different game forms were compared using root mean square error (RMSE) and the probability distribution. RMSE is calculated as (22):

$$RMSE = \sqrt{\frac{1}{n}\sum_{i=1}^{n}(y_i - \hat{y}_i)^2} \tag{22}$$

where $y_i$ represents the actual strategy, and $\hat{y}_i$ denotes the predicted strategy. A smaller RMSE signifies a lower error. The RMSE results are displayed in Fig. 9.

Furthermore, in the context of correctly predicted strategy, the probability distribution of correctly predicted strategies that were identified as Nash equilibrium states was subjected to statistical analysis, as illustrated in Fig. 9.

A smaller RMSE indicates a smaller discrepancy between the predicted outcomes of the model and the actual observations, thereby signifying an enhanced accuracy of the model's predictions. The probability distribution for the situation where the straight-going vehicle B moves first is more concentrated and approaches 1, with the smallest RMSE (0.409) observed, reaching an accuracy of 83.3%. This indicates a heightened level of confidence in predicting the outcome of this game. The concentrated distribution implies a lower probability of multiple strategies simultaneously being Nash equilibrium states, which in turn improves the accuracy of strategy selection.

As a result, the prediction results where the straight-going vehicle moves first in the game demonstrate higher accuracy. **The answer to "When the AV should disclose its intention?" is that the AV should disclose its intention information in advance.** The calibrated results are presented in TABLE II.

## V. RESULTS AND DISCUSSIONS

### A. The Standard Form of a Signaling Game

A standard structure of a signaling game is illustrated in Fig. 10. The game involves two players: the sender ($B$) and the receiver ($A$). $B$ has two possible types, $B_1$ and $B_2$, and can send one of two signals, $S_1$ or $S_2$. After receiving the signal, $A$ can take one of two actions, $A_1$ or $A_2$. The sequence of the game is as follows:

1) Nature ($\mathcal{N}$) moves first, assigning $B$'s type as $B_1$ with probability $p$ and as $B_2$ with probability $1-p$.

2) $B$ chooses to send either $S_1$ or $S_2$ according to its type.

3) $A$ receives the signal and selects an action, either $A_1$ or $A_2$.

4) There are a total of eight possible strategy combinations (denoted as $o_{ij}^k$. $i$ and $j$ still represent the strategies chosen by $A$ and $B$, respectively, while $k$ denotes $B$'s information disclosure strategy. Here $i, j, k \in \{1, 2\}$).

A crucial aspect of a signaling game is that $A$ forms a belief about $B$'s state based on the received signal. However, as indicated by the green nodes in Fig. 10, the information disclosed by $B$ is not necessarily truthful, meaning deception may occur.

### B. Process of EHMI Information Disclosure

We studied the EHMI information disclosure process in unprotected left-turn scenarios. It is assumed that the left-turning vehicle is a HV, and the straight-going vehicle is an AV with the functionality of EHMI.

Compared to the standard Stackelberg game, the signaling game not only incorporates the impact of action sequence but also introduces a signaling stage, leading to potential information asymmetry. As illustrated in Fig. 10, among the players, the leader corresponds to the sender, while the follower corresponds to the receiver. The signaling stage refers to EHMI information disclosure.

Additionally, based on real-world interaction scenarios, we have made certain modifications to the standard form of the signaling game. The main difference is that the sender ($B$) chooses the strategy independently, rather than it being assigned by nature ($\mathcal{N}$) with a certain probability.

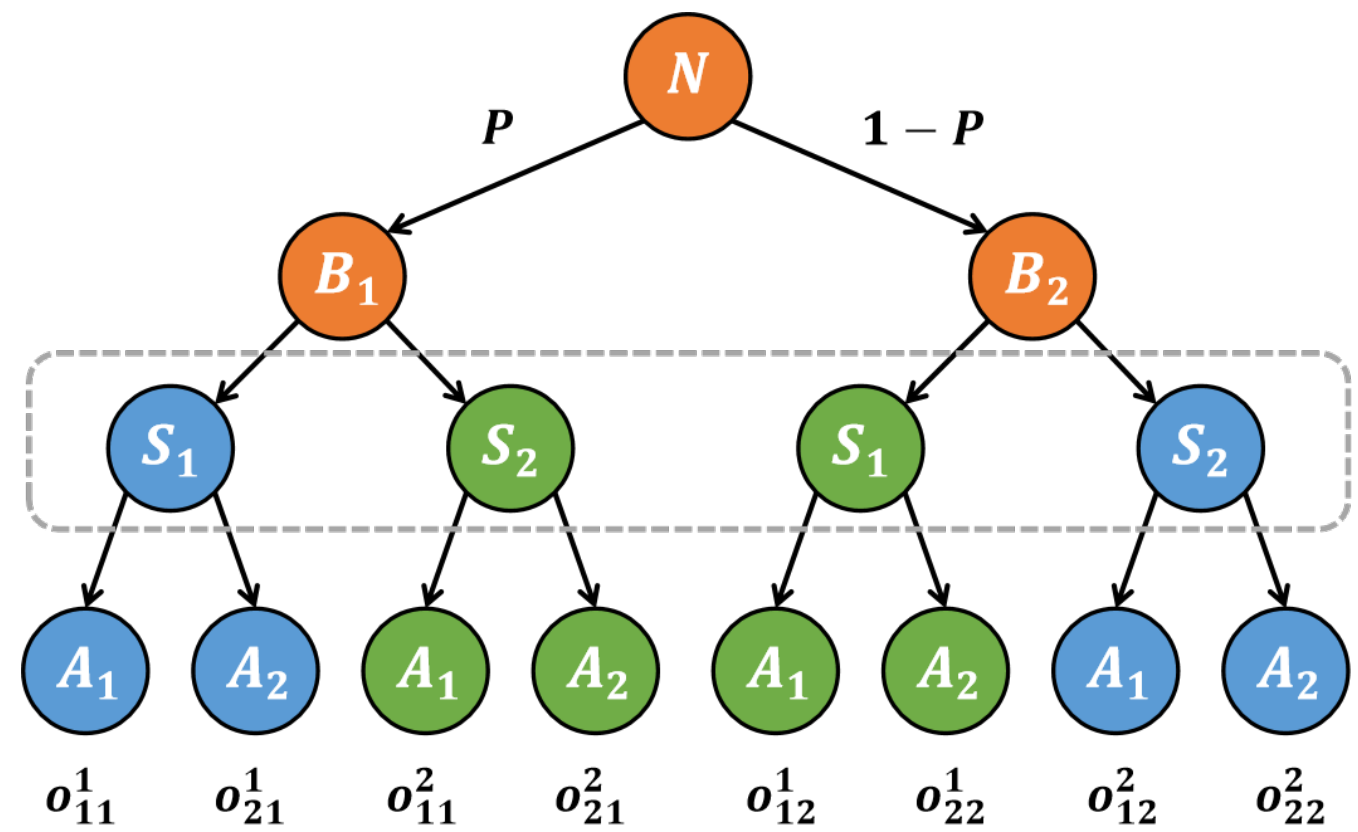

**Fig. 10.** A standard structure of a signaling game. Signaling games add an information disclosure stage between the acting steps of a Stackelberg game (nodes within the gray dashed box). Both truthful and deceptive information disclosures inevitably exist in a signaling game. Blue nodes denote truthful disclosure, and green nodes denote deceptive disclosure.

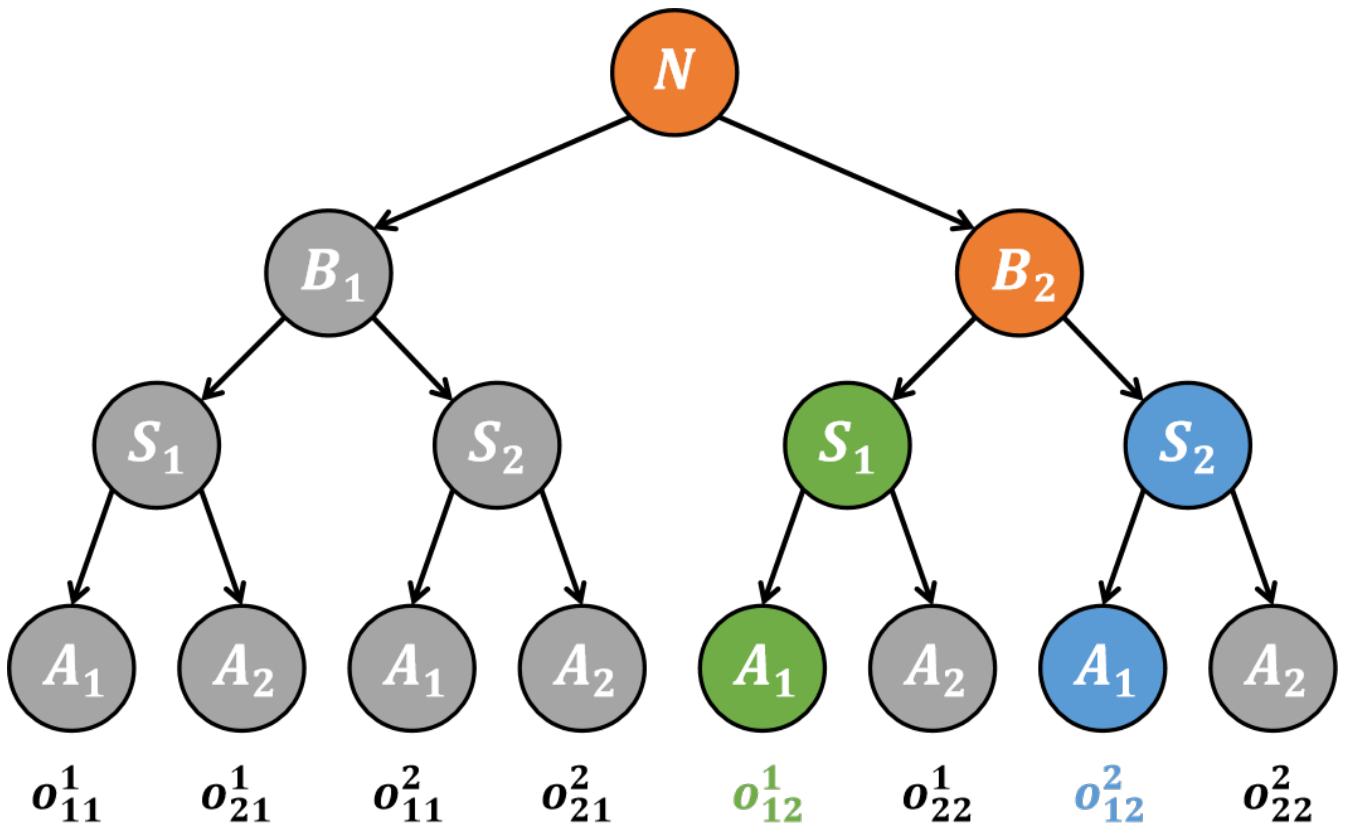

**Fig. 11.** In the signaling game, there are two ways to achieve $o_{12}$: Disclosing truthful information ($o_{12}^{2}$) and disclosing deceptive information ($o_{12}^{1}$).

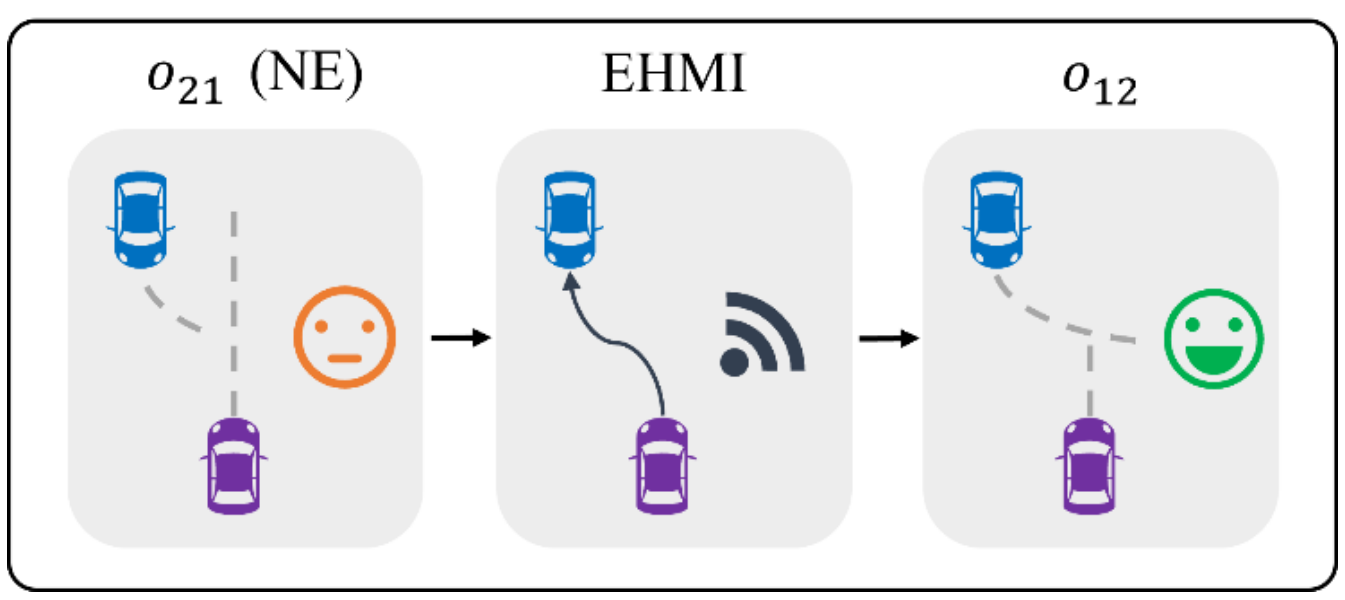

**Fig. 12.** The process of EHMI information disclosure. According to the game model, AV changes the interaction strategy from $o_{21}$ to the expected strategy $o_{12}$ using EHMI. In this way, the expected payoff is maximized.

From the perspective of the receiver ($A$), due to incomplete information, the perceived game structure differs from $B$'s when it is assumed that $A$ believes all disclosed information from $B$ is truthful. In this case, $A$'s view of the game excludes the green parts in Fig. 10, representing deceptive disclosures. Consequently, it is possible that $A$ believes the final strategy chosen is $o_{11}^{1}$, while the executed strategy is $o_{12}^{1}$ in reality.

Specifically, initially, the AV calculates the probabilities of potential strategies as Nash equilibrium states, based on the initial positional and motion information. The straight-going vehicle moves first in the game, so the absence of EHMI results in the subsequent interaction following strategy $o_{21}$ (in Fig. 6). But the AV can implement strategy $o_{12}$ by disclosing yielding information via EHMI (see Fig. 11 and Fig. 12).

The HV, acting later, will choose the strategy with the highest payoff in the game model upon receiving EHMI information. Therefore, the HV's chosen strategy does not necessarily always oppose the AV's disclosed strategy, allowing for the possibility of benevolent deception.

### *C. Impacts of EHMI Information Disclosure*

Firstly, we solely consider the impact of the expected strategy without considering the truthfulness of the information. This refers to the situation where only the true information is disclosed (considering only the blue nodes).

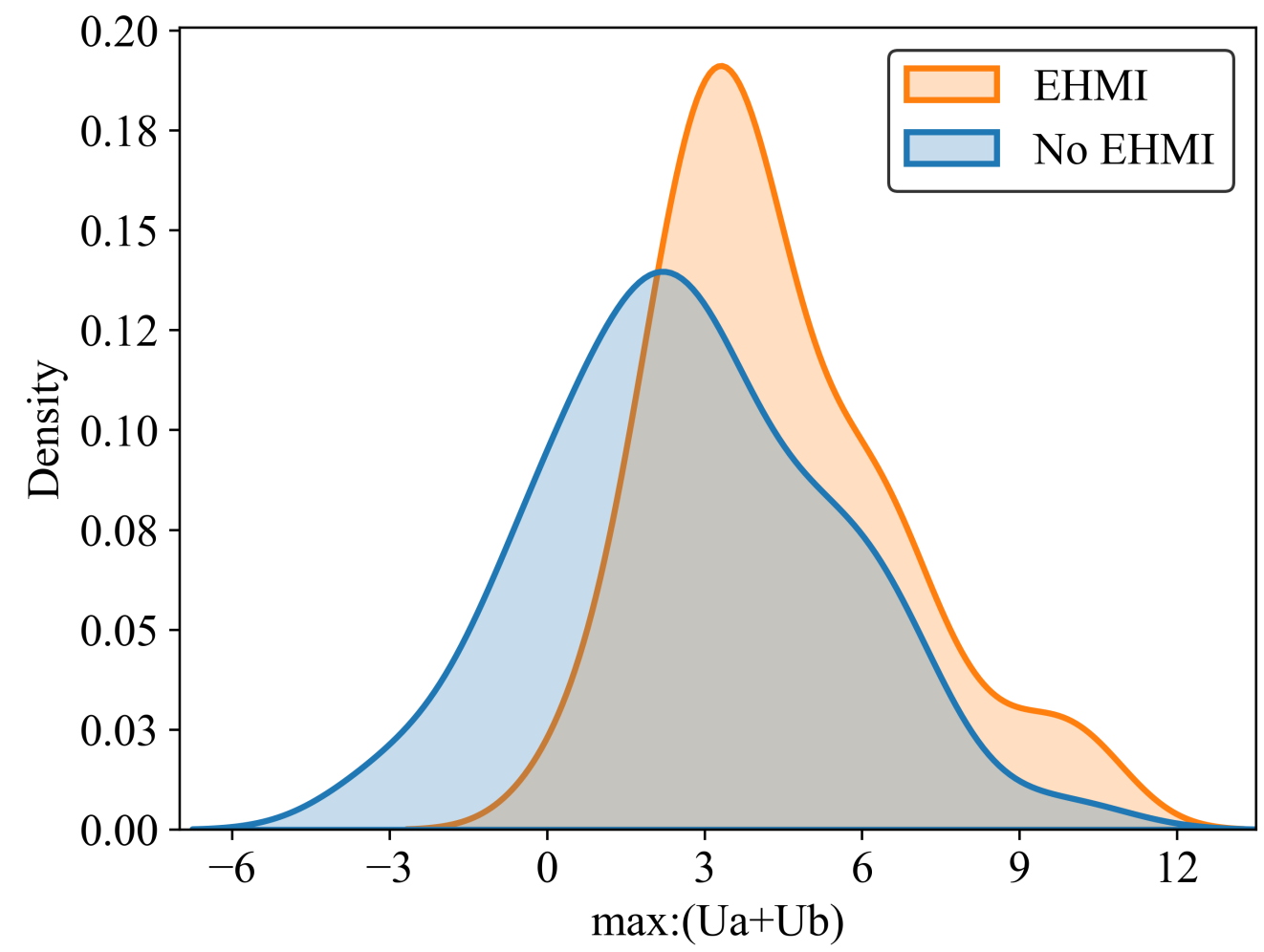

**Fig. 13.** Distribution of the total payoff for both players, which can be increased by EHMI.

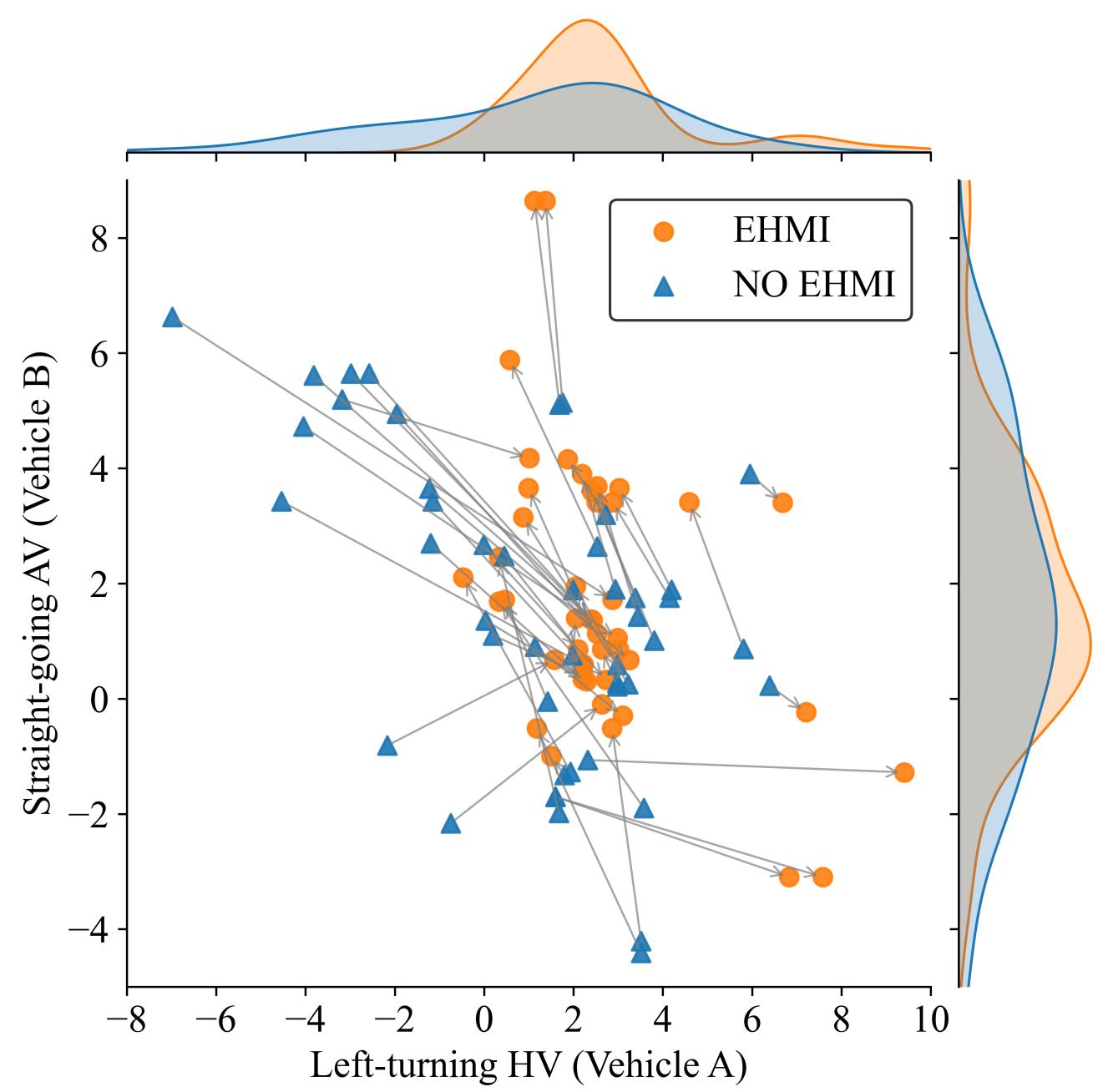

**Fig. 14.** Distribution of each players' payoff, which can be enhanced by EHMI.

Assuming the Nash equilibrium state is $o_{21}$, if we prioritize the maximum total payoff for both players with strategy $o_{12}$, the strategy selection can be modified by disclosing such information through EHMI. We extracted games where the Nash equilibrium states do not maximize the total payoff from the dataset of 484 instances of interactions. Among them, there are 47 instances (9.7%) in which optimizing the game through EHMI can maximize the total payoff. The distribution is illustrated in Fig. 13. The EHMI information disclosure strategy results in a significant rightward shift in the distribution of payoffs for the corresponding driving preferences, representing a noticeable increase in payoff. The average payoffs of the 47 instances were also shown in TABLE III.

TABLE III
THE AVERAGE CHANGES IN PAYOFFS WHEN EHMI IS APPLIED

| | No EHMI | EHMI | Increase |
|---|---|---|---|
| $U^A + U^B$ | 2.67 | 4.39 | 1.72 |
| $U^A$ | 1.13 | 2.63 | 1.51 |
| $U^B$ | 1.54 | 1.75 | 0.22 |

TABLE IV
THE AVERAGE CHANGES IN PAYOFFS FOR EACH PLAYER

| Types | $\Delta U^A$ | $\Delta U^B$ | Occurrences |
|---|---|---|---|
| $U^A\uparrow, U^B\downarrow$ | 4.32 | -2.29 | 21 |
| $U^A\downarrow, U^B\uparrow$ | -1.23 | 2.43 | 22 |
| $U^A\uparrow, U^B\uparrow$ | 1.81 | 1.20 | 4 |

Further analysis of the players' individual average payoffs was made. Both players experience an increase in payoff (shown in TABLE III), and the distribution is more concentrated compared to the scenario without EHMI (shown in Fig. 14). The average changes in the payoffs under three types are shown in TABLE IV. It can be observed that EHMI can significantly increase the lower-rewarded player's payoff by slightly reducing the higher-rewarded player's payoff. One unit of decrease in one's payoff can lead to nearly a doubled increase in the other's payoff, thereby improving the overall quality of the interaction. Besides, when both players have low payoffs initially, EHMI can enhance the payoffs for both players, which occurs for 4 times out of 47 cases.

The increase in payoff for left-turning HV (vehicle A) is more pronounced compared to the smaller increase for straight-going AV (vehicle B). Besides, the straight-going AV will choose to sacrifice its own payoff more aggressively to increase the left-turning HV's payoff, while it will be more conservative in sacrificing the HV's payoff. This emphasizes more payoff of the left-turning HV, enhancing the overall payoff compared to the scenario without information disclosure. In this way, EHMI can transform the non-cooperative game into a cooperative one. **So, the answer to "Whether the AV should disclose its intention?" is that the AV should disclose its intention information.**

*D. Impacts of Deception*

1) **Process of Benevolent Deception**

The premise for considering benevolent deception is that the AV has already determined an expected strategy.

Once the expected strategy is determined, if the AV uses EHMI to disclose information and achieves this expected strategy, it is regarded as a successful disclosure. If the AV's disclosed intent diverges from its behavior in the expected strategy, it is considered deceptive. If EHMI information disclosure meets both criteria, it is termed a successful deception.

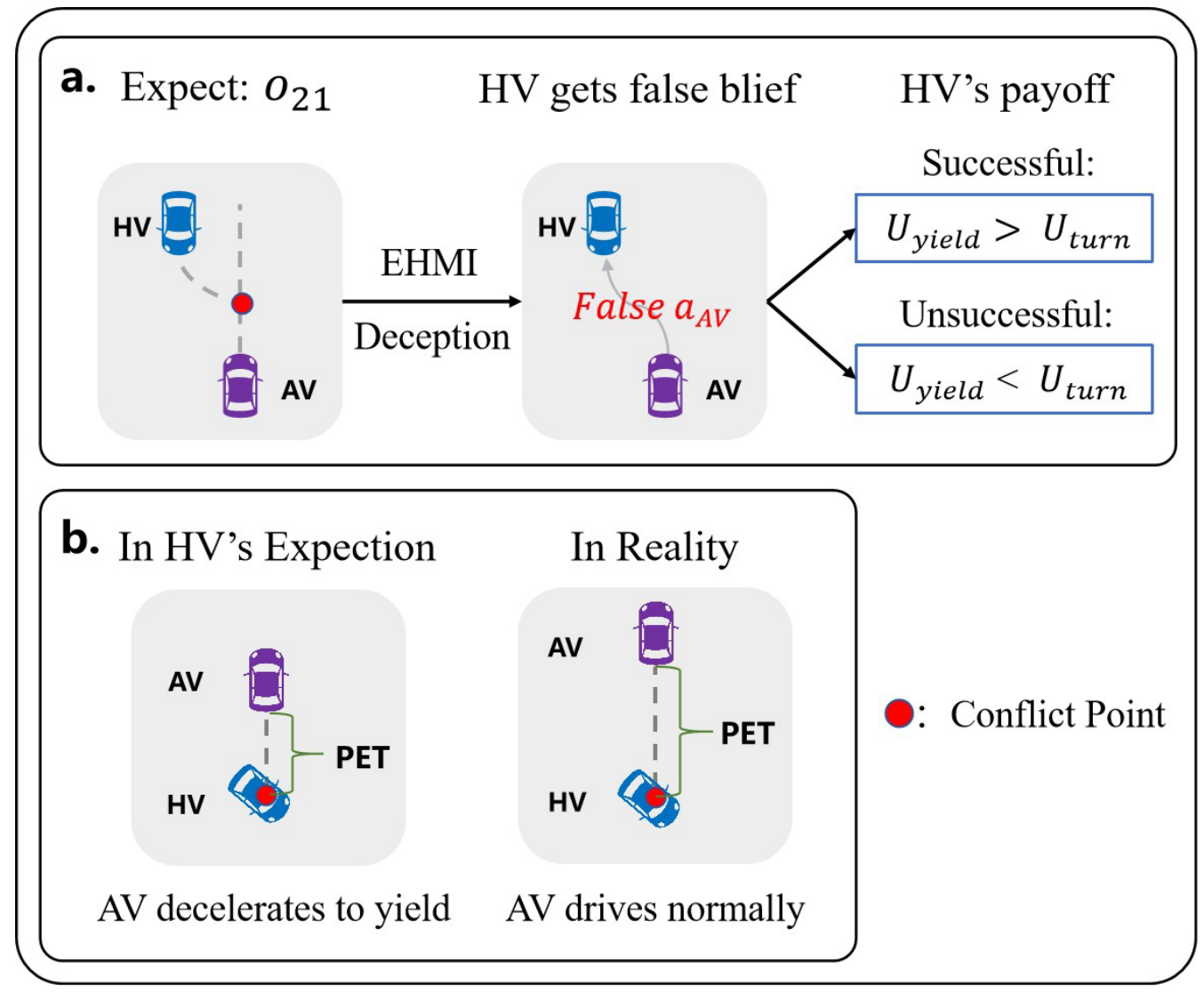

**Fig. 15.** a), Process and necessary conditions of deception. b), Successful benevolent deception can rise PET.

It is imperative to avoid unsuccessful attempts at deception. Similarly, if we assume that the anticipated strategy is $o_{21}$, a successful example of benevolent deception can be illustrated in Fig. 15(a):

a) The AV intends to pass through the conflict area first but discloses false intent information. This gives the HV a false belief. Specifically, the HV will then assumes the AV's acceleration corresponds to the average yielding acceleration.

b) If, under the false belief, the HV still considers yielding to have a higher payoff and chooses to yield, this constitutes a successful deception.

c) Since the AV doesn't actually intend to yield but chooses to pass through the conflict area first, the actual interaction will align with the expected strategy $o_{21}$.

Post-Encroachment Time (PET) is the time difference between the two vehicles passing the conflict point. Influenced by the false belief that the AV will decelerate to yield, the HV will increase its own deceleration to pass the conflict point after the AV (see Fig. 15(b)). And if the expected strategy is $o_{12}$, successful deception will cause the HV to increase its acceleration and pass through the conflict point first.

When the AV's driving behavior remains unchanged, the actual PET will increase, making interactions with aggressive human drivers safer.

2) **Impacts of Benevolent Deception**

We consider the average acceleration of the AV when passing first or later in the dataset as the belief given to the HV under deception with different expected strategies. The expected strategy of $o_{21}$ results in 28 successful deceptions. On the other hand, $o_{12}$ leads to 12. The two strategies together account for 8.3% of the total interactions.

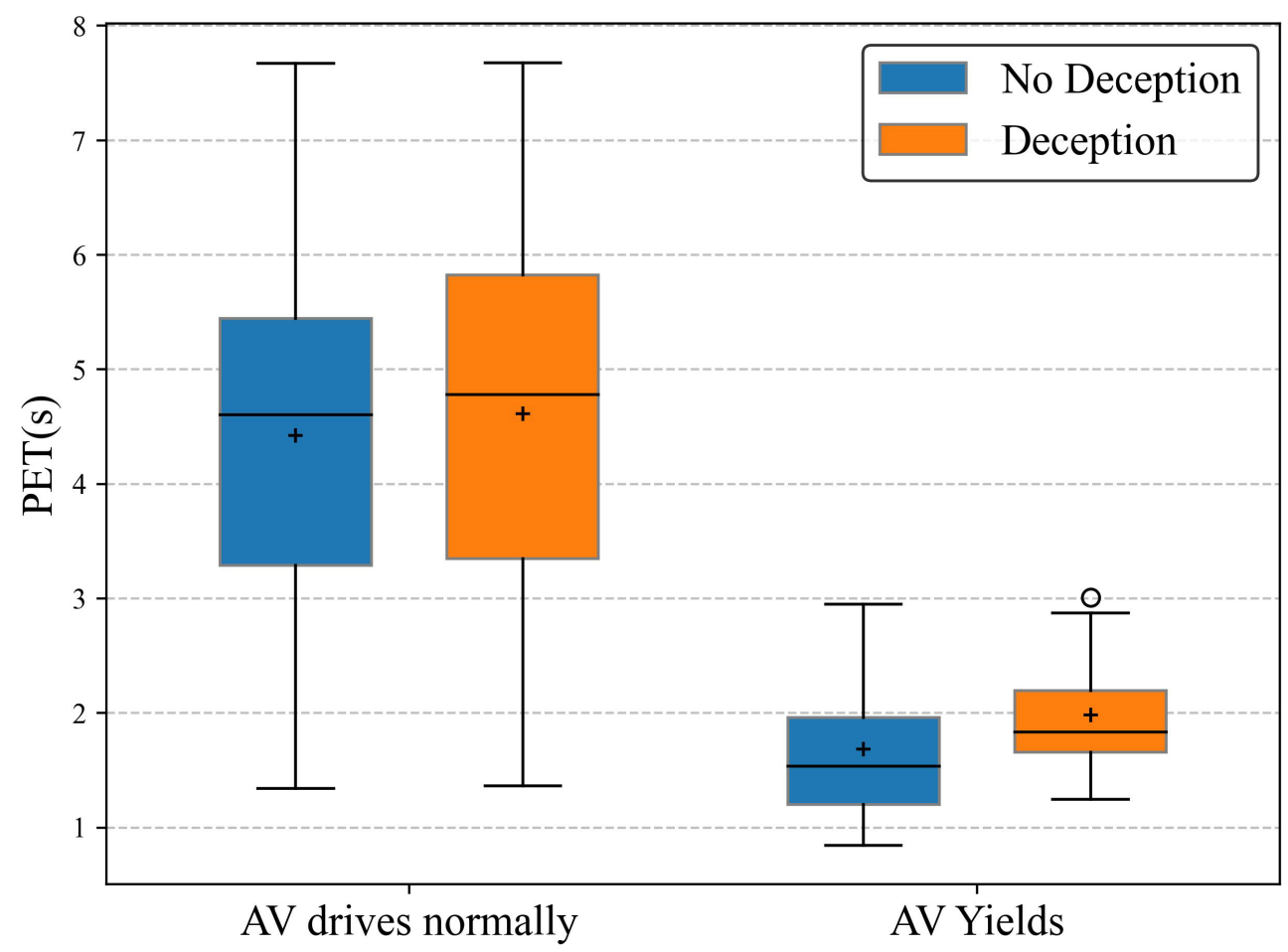

**Fig. 16.** Changes of PET when deception is applied. In both two cases, PET can be increased by successful deception.

TABLE V
PET(<3S) OF EXPECTED STRATEGY RESPECTIVELY

| Expected Strategy | No Deception | Deception |
|---|---|---|
| Straight-going AV First | 2.12 s | 2.23 s |
| Left-turning HV First | 1.68 s | 1.98 s |

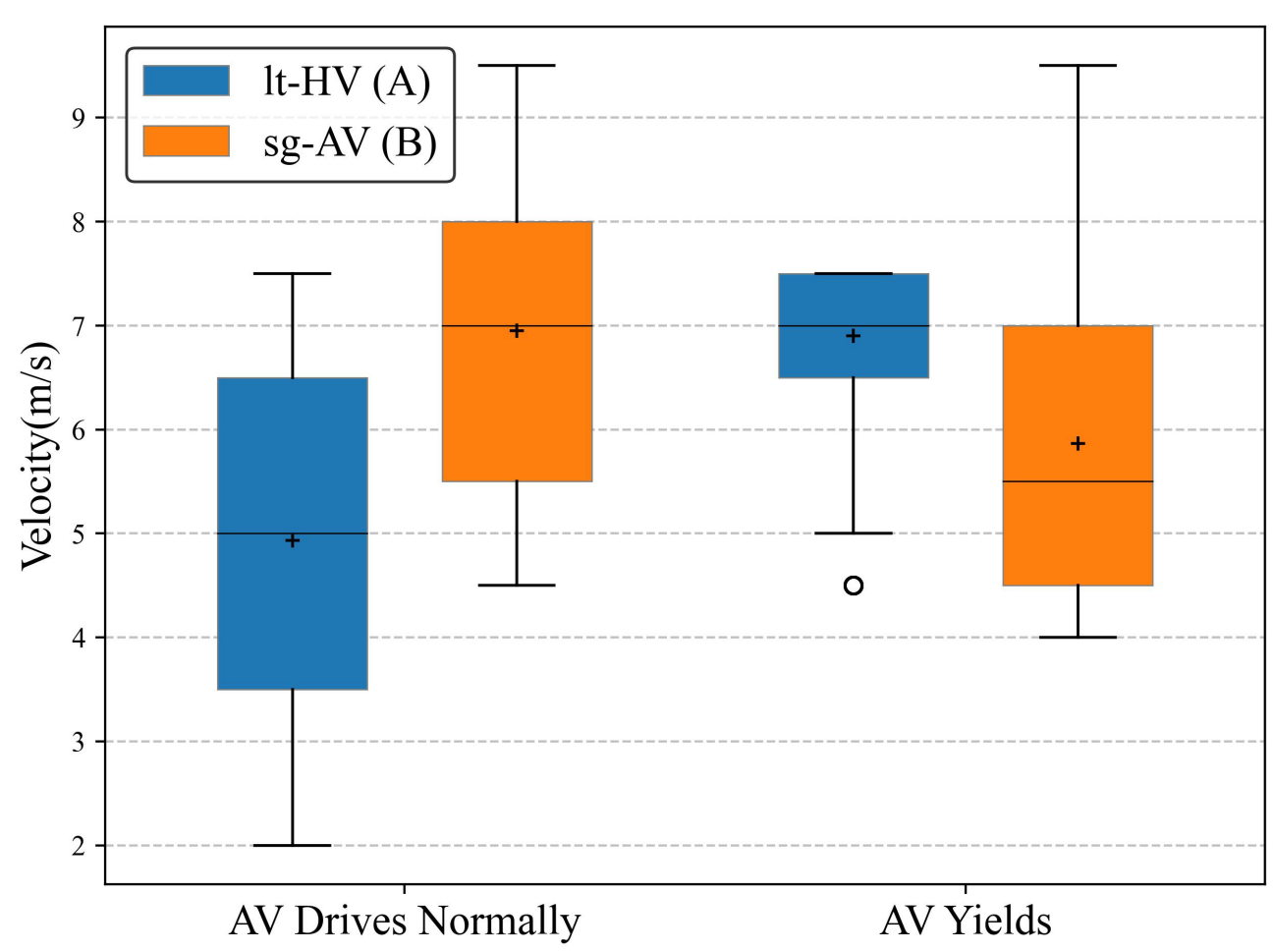

**Fig. 17.** The distribution of velocity under two expected strategies. Faster velocity is associated with a higher expectation of passing through first.

In our previous preliminary study [38], we calculated these successful deceptions using simple kinematic equations. The specific interaction trajectories may not correspond to reality. Fig. 16 illustrates the changes in PET. Table V shows the increase in PET before and after successful deception in dangerous interactions where PET < 3 s. When the expected strategy is for the straight-going AV to pass through the intersection first, the corresponding interaction is more dangerous, and successful deception significantly enhances safety.

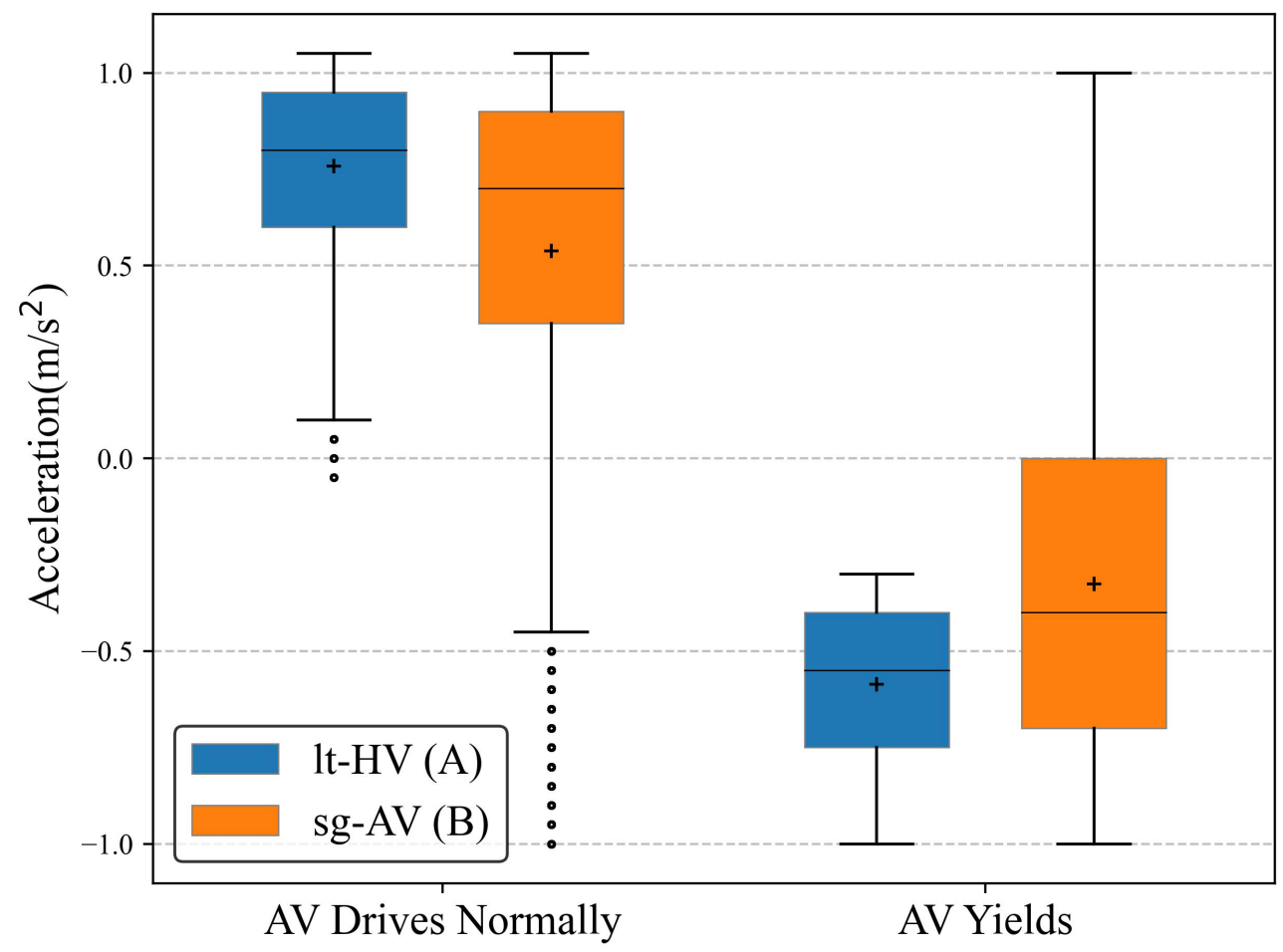

**Fig. 18.** The distribution of acceleration under two expected strategies. Higher acceleration leads to passing through later.

*E. Simulation Validation*

Based on the above analysis of EHMI information disclosure strategies, we conducted simulation experiments to further verify the necessity and feasibility of benevolent deception based on the expected strategy.

To consider the impact of successful deception in more general scenarios, we kept the map unchanged and first fixed the starting points, endpoints, and heading angles of the two vehicles. Then, based on the distribution of speeds and accelerations of the actual trajectories, we traversed the initial states ($v$, $a$) of the two vehicles. Using the established framework, we calculated whether successful deception could be achieved for each set of initial states. The initial state distributions of successful benevolent deception are illustrated in Fig. 17 and Fig. 18.

The velocity distribution (in Fig. 17) shows that the vehicle with the higher velocity expects to pass through the conflict point first, which aligns with intuition and previous analysis. However, the acceleration distribution (in Fig. 18) reveals a counterintuitive finding: the vehicle with higher acceleration expects to pass later. Theoretically, this could lead to dangerous conflicts. But this phenomenon just illustrates the role of benevolent deception, where HV's acceleration is influenced by false beliefs, causing HV to have higher acceleration when it expects to pass first and lower acceleration when it expects to pass later, thereby enhancing safety. Additionally, it can be observed that when AV expects to pass first, both vehicles tend to have positive acceleration, whereas when AV expects to pass later, the acceleration is mostly negative.

For the two expected strategies, we selected two typical initial states for trajectory simulation to further verify the effect of benevolent deception on improving interaction safety.

First, we planned the trajectories for both vehicles. Using the starting points, endpoints, and conflict points from the original data, we fitted approximate trajectories using B-spline

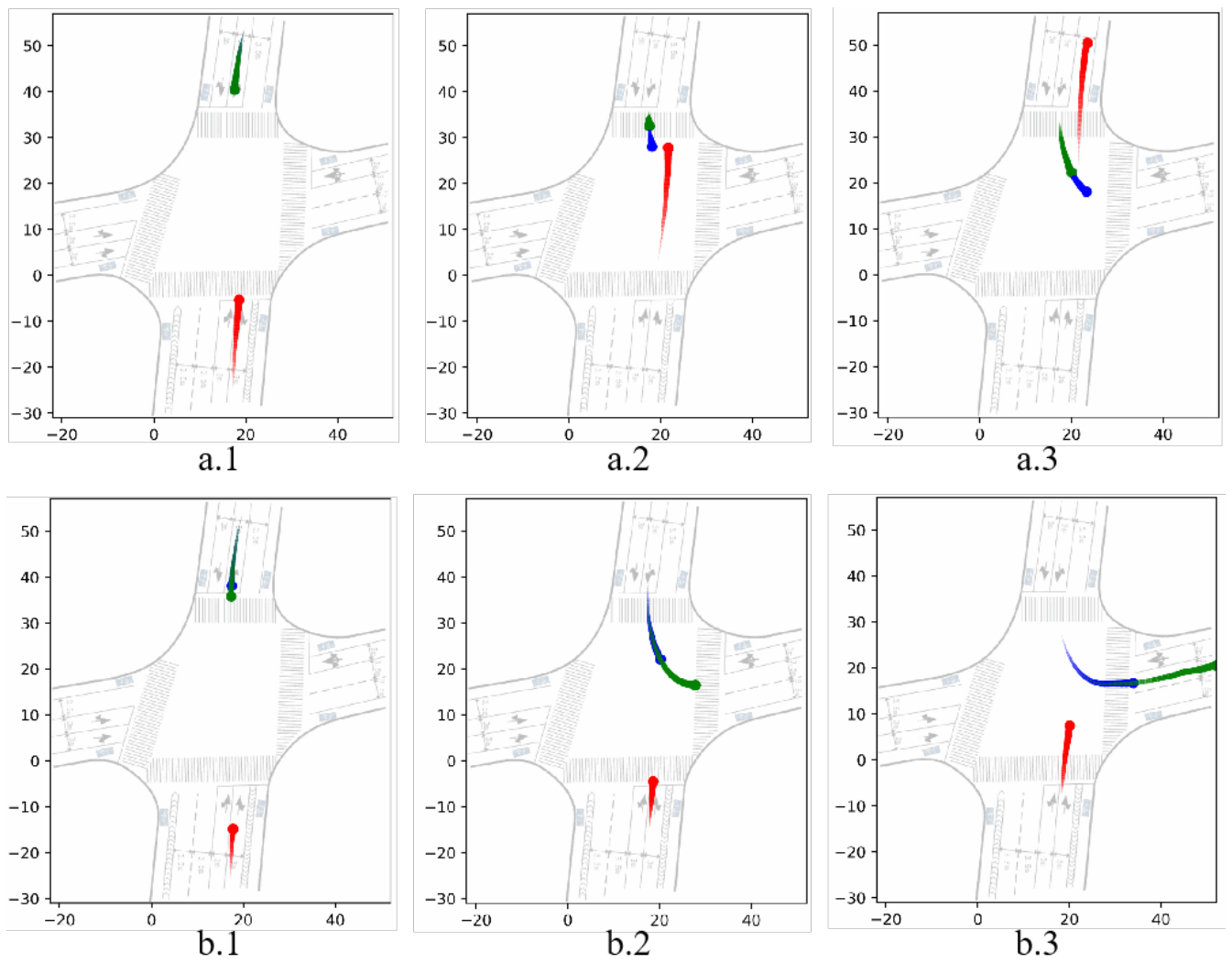

**Fig. 19.** Simulation results of two expected strategy. Red for AV, blue for HV without deception, green for HV with deception. a), AV drives normally and passes through the conflict zone first. PET increased from 2.7 s to 3.4 s. b), AV yields and passes through the conflict zone later. PET increased from 2.8 s to 3.9 s. The detailed process can be viewed in the video attached.

curves.

For real-world applications, some advanced control algorithms have been proposed. Ding et al. [39] proposed an adaptive memory event-triggered output feedback finite-time lane-keeping control strategy. Zhang et al. [40] introduced an event-triggered finite-time adaptive sliding mode coordination control method. However, we selected the rear wheel feedback control method primarily because it simplifies the validation process while ensuring reliable simulation results. Inputs are updated every 0.1 s, with the outputs being the position coordinates, heading angle, and speed.

The driving strategy for both vehicles is set to travel from the starting point to the stop line at the entrance using the current acceleration. Once one vehicle passes the conflict point, both vehicles accelerate through the intersection at 2 m/s². In the case of successful deception, after the stop line, the left-turning HV will travel using the average acceleration.

When the desired strategy is for the straight-going vehicle to pass first, the average yielding acceleration for the left-turning vehicle is set to -1.5 m/s² in the non-deceptive case and -2.5 m/s² in the successful deception case. When the desired strategy is for the straight-going vehicle to pass later, the average acceleration for the left-turning vehicle is set to -0.5 m/s² in the non-deceptive case and 0 m/s² in the successful deception case.

The trajectory simulation results are shown in Fig. 19. It can be observed that in cases where successful deception is possible, not choosing to deceive can lead to dangerous situations with PET < 3 s. Successful deception significantly increases the PET, thereby enhancing safety. Additionally, in the traversal results, the total proportion of successful deception is 14.05%, which is slightly higher than the actual data. Among these, the percentages for the two expected strategies are 8.48% and 5.57%, respectively. While this may not be a significant percentage, it is a notable factor. Therefore, carefully considering the need for benevolent deception is quite urgent.

### *F. Empirical Simulation based on VR*

Based on the above analysis of EHMI information disclosure strategies, we conducted simulation experiments to further verify the necessity and feasibility of benevolent deception based on the expected strategy. We used the experimental platform (TransCAVE) at the College of Transportation Engineering, Tongji University. The ego vehicle driven by participants can interact with background vehicles generated by the platform simultaneously.

We selected the experimental scene from the Tongji test site, as shown in Fig. 20. The participants operate the HV, making a left turn at the intersection along the blue route, while the algorithm-controlled AV equipped with EHMI travels straight through the intersection along the red route. Considering that EHMI may increase the driver's visual and cognitive load, our EHMI design was deliberately simplified and tuned to ensure perceptual clarity. Specifically, we use a red arrow to represent the AV disclosing "I want to proceed first" and a green arrow to represent the AV disclosing "I want to yield."

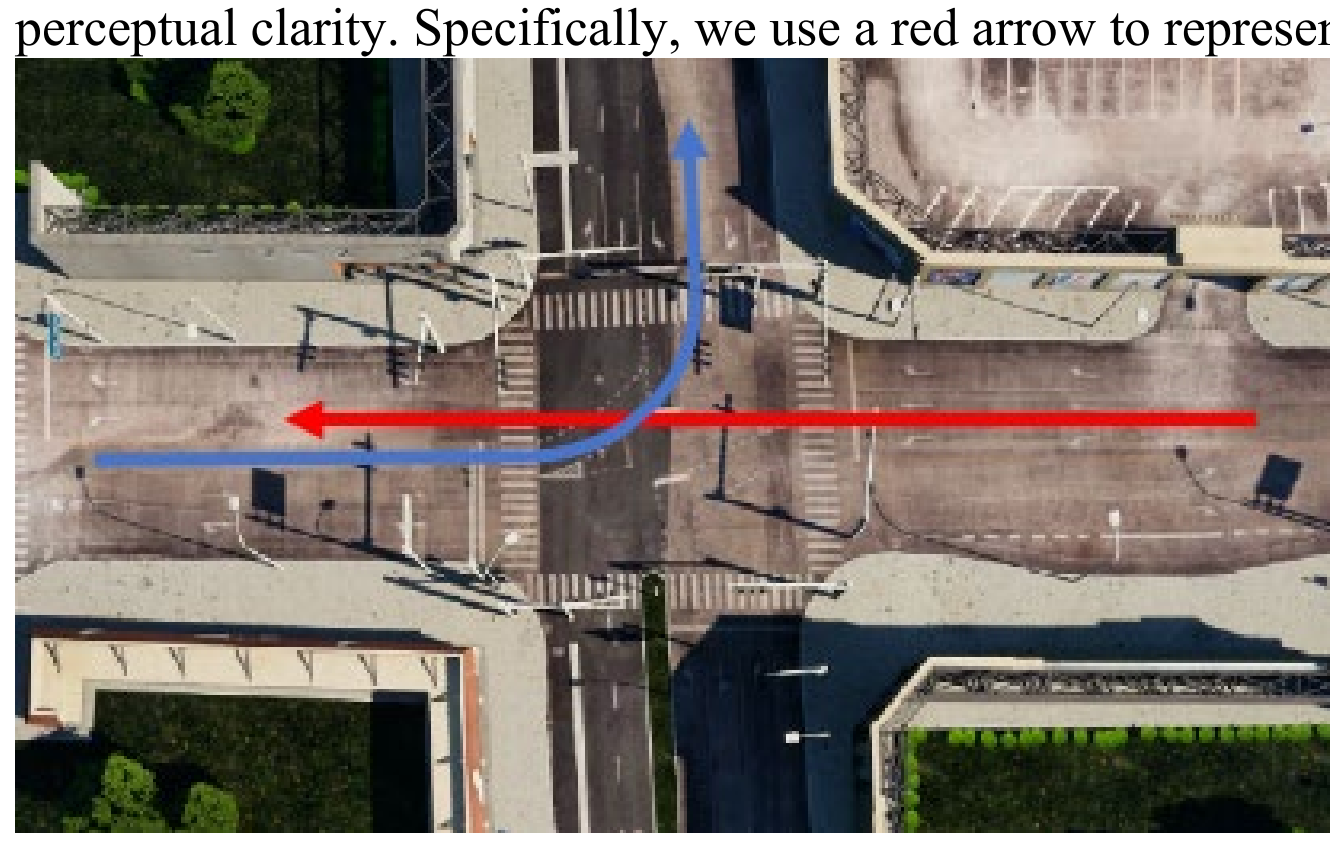

**Fig. 20.** Experimental scenario and driving routes. The blue and red lines represent the driving routes of the HV and AV, respectively.

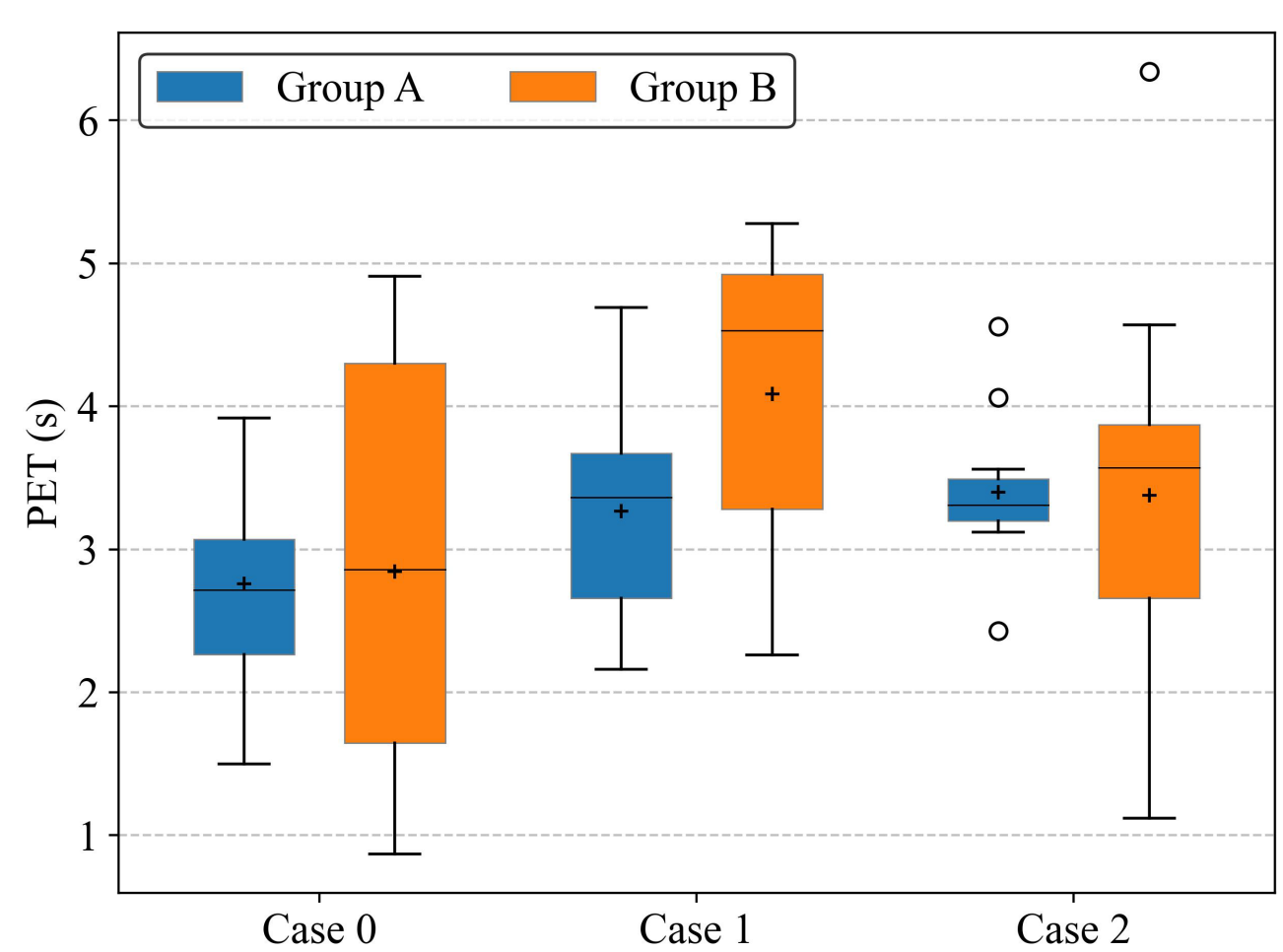

**Fig. 21.** Distributions of PET in unprotected left-turn scenarios. Compared to not using EHMI (Case 0), the PET shows a significant increase when EHMI is used.

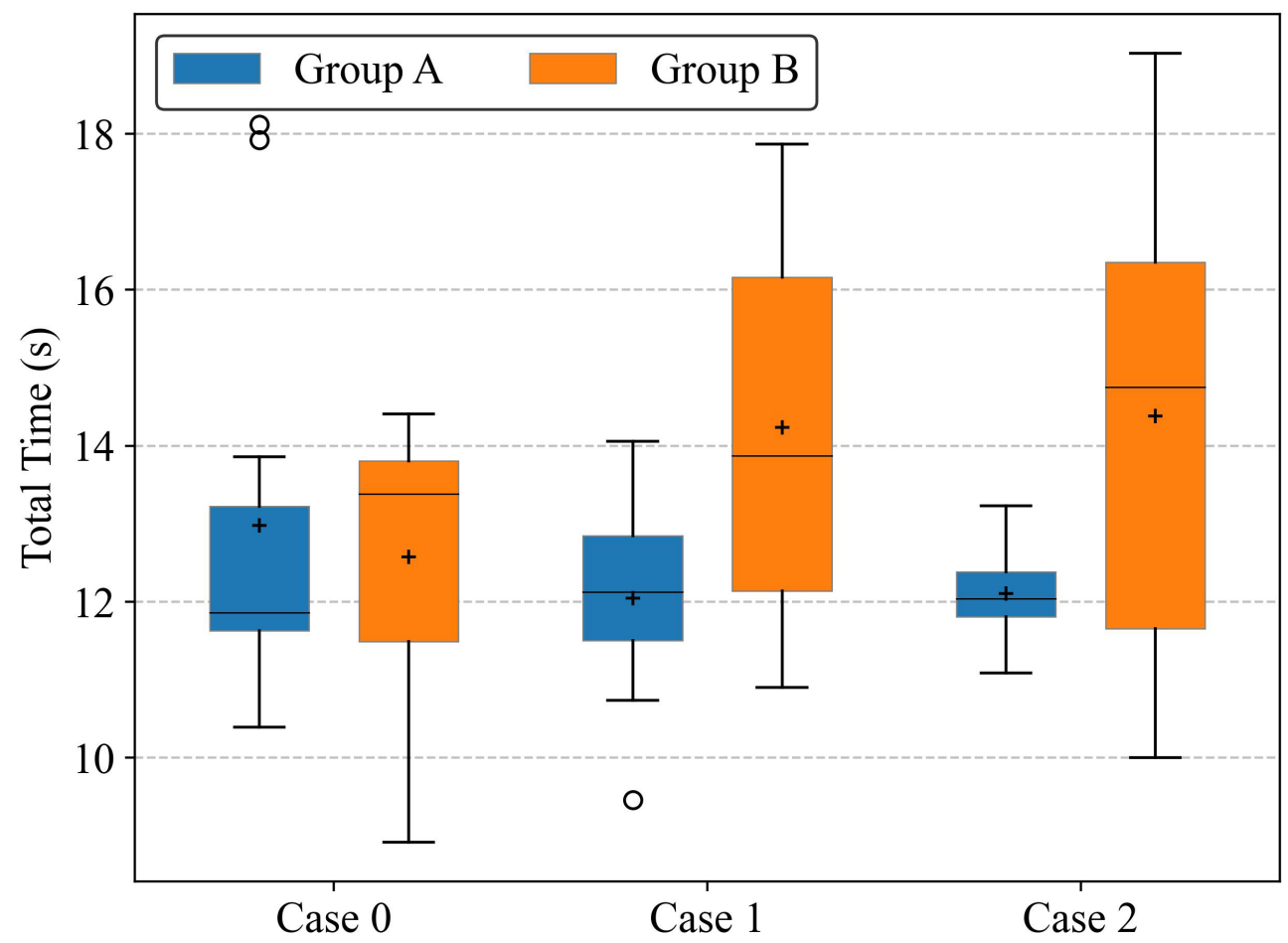

**Fig. 22.** Distributions of the total time taken by both vehicles to pass through the conflict point. The total time in the high HV trust scenario (Group A) is significantly lower than in the low HV trust scenario (Group B).

TABLE VI
QUESTIONNAIRE RESULTS OF HUMAN TRUST LEVELS

| Group | Case | Mean | SD |
|---|---|---|---|
| A | 1 | 8.250 | 1.215 |
| A | 2 | 8.083 | 1.240 |
| B | 1 | 5.167 | 1.992 |
| B | 2 | 4.583 | 1.564 |

Each participant will first drive freely in the scene for 10 minutes to be familiar with the environment before conducting 3 cases. The setup of the experimental cases are as follows:

- Case 0: No EHMI.
- Case 1: EHMI always discloses true information.
- Case 2: EHMI discloses information according to the signaling game.

In our experiment, a total of 24 participants (14 males: mean age = 23.93, range = 21–27 and 10 females: mean age = 22.9, range = 21–25) were recruited. All of them were licensed drivers (with 3–9 years of driving experience), and most of them had some AV and VR experience.

We divided the 24 participants into two groups equally. Group A remained uninformed about EHMI information disclosure prior to the three cases, symbolizing the current state of relatively high trust in EHMI. We informed Group B in advance that EHMI might disclose false information, indicating a potential future state with reduced trust levels.

Due to the constraints of platform simulation with the VR equipment, the experimental update frequency was set to 0.5 seconds. The EHMI update frequency was set to 2 seconds, ensuring that each piece of information remained visible and unchanged for at least 2 seconds. This setting was designed to maintain a balance between perceptibility and cognitive load in human–vehicle interaction.

1) **Experimental Results**

Fig. 21 shows the distributions of PET in different cases. When human drivers were unaware that the EHMI might be deceptive (Group A), using EHMI significantly improved safety. Compared to the scenario without EHMI (Case 0: PET = 2.80 s):

- Disclosing fully truthful information with EHMI increased PET by 18.5% (Case 1: PET = 3.27 s).
- Disclosing information that may not be entirely truthful increased PET by 23.3% (Case 2: PET = 3.40 s).

Both approaches led to better safety outcomes.

Furthermore, for a subset of participants (6 out of 12, 50%) in Group A, the PET in Case 2 (3.59s) was higher than in Case 1 (2.92s), with a 22.9% increase. This indicates that in specific dangerous interactions (PET < 3 s), it can significantly enhance safety.

2) **The Impact of Reduced Trust Levels**

As shown in Fig. 21, the PETs of the two groups are similar in Case 2. However, in Case 1, Group B has a significantly higher PET than Group A. The EHMI information disclosure framework was designed under the assumption of high HV trust. Theoretically speaking, it should perform worse in terms of safety when the trust level is low. However, it turns out that the observed PET is higher (better safety performance) when the trust level is low as shown in the experiment, which contradicts the initial assumption.

In addition to safety, we also considered efficiency. And we studied the total time it takes for the two vehicles to pass through the conflict point. The distributions under different cases are shown in Fig. 22. When the participants were informed in advance that deceptive information might be disclosed (Group B), the interaction efficiency significantly decreased regardless of the EHMI information disclosure strategy.

Moreover, in this situation, using EHMI results in even lower efficiency compared to not using EHMI (Case 0). Thus, we believe that a decrease in HV trust level is detrimental to the efficiency of the interaction.

3) **The Impact on Human Trust Levels**

After each experimental case, participants were asked to rate their trust level in the EHMI. The rating scale ranged from 0 to 10, with higher scores indicating greater trust. The subjective feedback (in TABLE VI) revealed an interesting phenomenon: regardless of whether deception occurred, explicitly informing participants about the possibility of deception (Group B) significantly reduced their trust compared to Group A, where no information was provided. Therefore, people are also unlikely to detect whether the interaction partner is deceiving them.

*G. Discussions of Benevolent Deception*

This study is positioned as an exploratory behavioral mechanism analysis based on specific hypotheses and interaction contexts. From a modeling standpoint, benevolent deception is shown to be both feasible and potentially beneficial under certain conditions. **However, as for the last question, "whether it should disclose the true intention or provide a false intention?" drawing from Asimov's Three Laws, different perspectives may arise when considering ethical and societal development aspects.**

Some argue that any form of deception undermines transparency and long-term public trust, even if short-term safety benefits exist. Our finding also provided preliminary evidence that such long-term trust effects could potentially lead to worse outcomes, such as reduced interaction efficiency. Others suggest that if deception benefits all parties—by improving safety or efficiency—may be ethically acceptable. And the long-term trust concerns may be resolved through alternative approaches.

Given the early stage of public acceptance and regulatory development in autonomous driving, we believe ethical strategies should avoid controversial or aggressive approaches. Besides, our aim is not to advocate the use of deception, but to explore its interaction effects under specific controlled conditions. This helps inform the future research and design of more transparent, adaptive, and socially acceptable EHMI systems.

## VI. Conclusion

In this study, we established a comprehensive framework for the EHMI information disclosure decision-making process in interactions between AV and HV, incorporating the possibility of benevolent deception for the first time. Through signaling game modeling, data calibration, and simulation experiments, we compared, analyzed, and explored the proposed framework from a mechanistic perspective. This approach effectively maximizes the benefits for both. And it is required to disclose the intention, preemptively disclose the intention, and authentically disclose the intention. The current findings should be interpreted as a theoretical exploration rather than definitive guidance for future design and ethical decision-making.

Future research needs to address several aspects: 1) Larger datasets from different locations and different calibration methods are essential to enhance the generalizability and credibility of the model. 2) Final validation through real vehicle testing is necessary to fully replicate all sensations, such as acceleration. 3) Different degrees of disclosed information such as frequency and details under the framework of signaling games warrant further investigation. 4) The current assumption is that human drivers consider the intentions disclosed by AVs to be genuine. If benevolent deception becomes widely known and adopted in the future, it is worthwhile to explore the impacts of human's trust, and further the deception between AVs.